\documentclass[12pt, draftclsnofoot, onecolumn]{IEEEtran}
\usepackage[usenames]{color}
\usepackage[dvips]{epsfig}
\usepackage{amssymb}
\usepackage{amsmath}
\usepackage{array,hhline,arydshln}
\usepackage{multirow}
\usepackage{graphicx}
\usepackage{latexsym}

\begin{document}
\title{Joint Beamforming Design for Multi-User Wireless Information and Power Transfer}
\author{Hyukmin Son, \IEEEmembership{Member,~IEEE} and Bruno Clerckx, \IEEEmembership{Member,~IEEE}
\thanks{This research was supported by Basic Science Research Program through the National Research Foundation of Korea (NRF) funded by the Ministry of Education, Science and Technology}
\thanks{H. Son and B. Clerckx
are with Communications and Signal Processing Group
Department of Electrical and Electronic Engineering
Imperial College London, South Kensington Campus
London SW7 2AZ, United Kingdom.(e-mail:
h.son@imperial.ac.uk and b.clerckx@imperial.ac.uk
)(Tel: +44-(0)20-7594-6234, Fax: +44-(0)20-7594-6302)
}}\maketitle
\begin{abstract}
In this paper, we propose a joint beamforming algorithm for a multiuser wireless information and power transfer (MU-WIPT) system that is compatible with the conventional multiuser multiple input multiple output (MU-MIMO) system. The proposed joint beamforming vectors are initialized using the well established MU-MIMO zero-forcing beamforming (ZFBF) and are further updated to maximize the total harvested energy of energy harvesting (EH) users and guarantee the signal to interference plus noise ratio (SINR) constraints of the co-scheduled information decoding (ID) users. When ID and EH users are simultaneously served by joint beamforming vectors, the harvested energy can be increased at the cost of an SINR loss for ID users. To characterize the SINR loss, the target SINR ratio $\mu$ is introduced as the target SINR (i.e., SINR constraint) normalized by the received SINR achievable with ZFBF. Based on that ratio, the sum rate and harvested energy obtained from the proposed algorithm are analyzed under perfect/imperfect channel state information at the transmitter (CSIT).
Through simulations and numerical results, we validate the derived analyses and demonstrate the EH and ID performance compared to both state of the art and conventional schemes.
\end{abstract}

\section{Introduction}
\label{intro}

In recent years, wireless power transmission has appeared as a promising technology to directly harvest energy from broadcasted RF signals \cite{Minhong}\cite{Ugur}. The interesting point is that all kinds of broadcast signal can be used to harvest energy, even though some broadcast signals are interference signals from the perspective of information decoding (ID). Hence, it is possible to transfer information and power to mobile devices simultaneously. The RF signal based energy harvesting system has drawn significant interest recently in wireless communication community, namely in simultaneous wireless information and power transfer (SWIPT) systems. In comparison with a near field wireless harvesting technique based on electromagnetic induction \cite{Andre2007}, the RF signal based energy harvesting can deliver power to mobile devices far from transmitters. This opportunity would allow battery-free devices, which can be free of connectors and have freedom of placement and mobility during charging and usage. Although RF signal based energy harvesting is typically suitable for low-power applications such as sensor networks or Internet of Things, it may be envisioned that power consuming applications could benefit from RF energy harvesting in the future if efficient dedicated wireless power transmission is implemented.

Previous works in SWIPT systems focused on the fundamental performance limits \cite{Lav}-\nocite{Grover}\nocite{Rui2}\nocite{Xun}\cite{Rui2011}.%
In \cite{Lav}, the fundamental tradeoff between EH and ID rates was studied in a point-to-point single-antenna additive white Gaussian noise (AWGN) channel from the information theoretic viewpoints. This work was then extended to frequency-selective AWGN channels considering a noisy coupled-inductor circuit in \cite{Grover}.
The authors in \cite{Rui2} researched opportunistic energy harvesting from the received unintended interference and/or intended signal in addition to decoding the information. Under a point-to-point flat-fading channel with time-varying interference, they derived the optimal ID/EH mode switching rules at the receiver to optimize the outage probability/ergodic capacity versus harvested energy trade-offs.
In \cite{Xun}, two practical receiver architectures, namely  the separated and integrated information and energy receivers, were proposed based on a dynamic power splitting (DPS) strategy. The rate-energy tradeoff for the two architectures are characterized from the rate-energy
region. SWIPT in multiple antenna system is then investigated in \cite{Rui2011}.
In that paper, the authors analyze the optimal rate-energy (R-E) tradeoff regions for two different cases, i.e. separated and co-located receiver scenarios in multiple input and multiple output (MIMO) broadcast channel. In addition, time switching and power splitting receivers are designed for the co-located receiver and they are characterized by their achievable R-E regions.

The beamforming for multiple receivers and/or transmitters were investigated based on perfect channel state information at transmitter (CSIT) in \cite{park2013}\cite{Jie2013}. In \cite{park2013}, the authors propose signal-to-leakage-and-energy ratio (SLER) maximization beamforming in a two user MIMO interference channel and identify the achievable R-E tradeoff region.
The work in \cite{park2013} has also been extended recently to the general K-user interference channel in \cite{park1}. In \cite{park2}, a geodesic beamformer design was found to be a suitable strategy in the two-user interference channel with perfect but partial CSIT. SWIPT for multiuser MISO is researched in \cite{Jie2013}. The authors in \cite{Jie2013} formulated the joint information and energy transmit beamforming problem as a non-convex quadratically constrained quadratic program (QCQP), for which the optimal solution is obtained by applying the technique of semidefinite relaxation (SDR). In particular, it is observed that the dedicated EH beamforming vectors are not necessary. Consequently, the optimal beamforming can be obtained by solving the SDR of the optimization problem.
The optimization problem can be solved using existing software, e.g. CVX \cite{CVX}, but it would lead to high complexity and cannot be realistically implemented in a practical system \cite{complexity}.
Furthermore, in a general multi-user MIMO (MU-MIMO) where scheduling and beamforming are performed, finding the optimal solutions of the optimization problem becomes intractable due to its non-convexity.

The problem of imperfect CSIT in SWIPT systems has been investigated in \cite{Zheng2012}\cite{Derrick2014}. The authors of these papers consider the deterministic imperfect CSI model to formulate and solve optimization problems regarding energy harvesting. However, it is still necessary to develop and analyze a \emph{practical}\footnote{The term of '\emph{practical}' means implementable from the viewpoint of low complexity and compatibility with conventional limited feedback based MU-MIMO systems. The details are discussed in Section \ref{art}, \ref{algorithm} and \ref{feedback}} joint beamforming for a limited feedback based SWIPT systems.

In this paper, to maintain compatibility with the conventional MU-MIMO system, the joint beamforming vectors are initialized using the well established MU-MIMO zero-forcing beamforming (ZFBF)\footnote{The conventional beamforming schemes for MU-MIMO can be adopted for the proposed joint beamforming scheme. Since ZFBF with semi-orthogonal user selection (SUS) is well known as a relatively simple and effective linear transmission technique in MU-MIMO system and is used in realistic communication system as LTE-A \cite{spencer2004}-\nocite{yoo2006}\cite{Lim2013}, ZFBF is utilized as the initial beamforming vectors in multi-user wireless information and power transfer system.} and are further updated to maximize the total harvested energy and guarantee the signal to interference plus noise ratio (SINR) constraints of the co-scheduled ID users in multiuser wireless information and power transfer (MU-WIPT).
For joint beamforming vector design, BS iteratively selects a beamforming vector\footnote{Initially, the beamforming vectors are constructed as ZFBF based on the selected ID users via SUS.}  dedicated to the specific ID user and steers it to an optimal direction to maximize the harvested energy while satisfying the SINR constraints of the selected ID users.

In the proposed joint beamforming algorithm, we introduce the EH gradient based criterion to select the beamforming vector among the candidate beamforming vectors dedicated to the selected ID users. To identify the optimal EH direction
to maximize the harvested energy via beamforming vector steering, it is demonstrated in \emph{Theorem 1} that the energy harvested from arbitrary beamforming vectors form a $\min(K_{EH},M)$-dimensional ellipsoid in the given EH users' channel space where $M$ is the number of transmit antennas and $K_{EH}$ is the number of EH users. It implies that an arbitrary beamforming vector should be steered along the geodesic line from the arbitrary beamforming vector towards the eigenvector corresponding to the largest eigenvalue for a given EH users' channel space.
At the end of each iteration in the proposed algorithm, the optimal direction and EH gradient are updated not to create additional interference to ID users who are at the edge of the given SINR constraints. By steering and updating the beamforming vectors, the SINR constraints are guaranteed and the total harvested energy is maximized in the proposed algorithm.

When ID and EH users are simultaneously served by joint beamforming vectors, the harvested energy can be increased at the cost of an SINR loss for ID users. To characterize the SINR loss, the target SINR ratio $\mu$ is defined as the target SINR (i.e., SINR constraint) normalized by the received SINR achievable with ZFBF.
Then, $1-\mu$ can be interpreted as the relative SINR loss (w.r.t. the maximum SINR achievable with ZFBF).
Based on the defined parameter $\mu$, we derive lower and upper bounds for the total harvested energy and sum rate under perfect/imperfect CSIT, respectively. By changing $\mu$, the joint beamforming algorithm trades harvested energy with sum rate and inversely. Furthermore, the asymptotic harvested energy and sum rate performance are analyzed for scenarios where the number of EH or ID users increases to infinity.


In Section \ref{system}, the system model for MU-WIPT is described and the complexity of the optimization problem in \cite{Jie2013} is discussed to motivate the necessity of the proposed joint beamforming algorithm. The proposed beamforming is then detailed in Section \ref{optimal_EH}. The achievable sum rate and harvested energy including asymptotic perspective are analyzed in Section \ref{analysis}. Simulations and numerical results are provided in Section \ref{simulation} in order to validate the performance of the proposed algorithm and the analyses.

For notations, we use lowercase lightface for scalar values, uppercase boldface for matrices,
lowercase boldface for vectors, and $(~\cdot)^H$ indicates the
conjugate transpose operator. Similarly, $(~\cdot)^{T}$ and
$(~\cdot)^{-1}$ indicate the transpose and inverse (or
pseudo-inverse) operators, respectively. $rank(~\cdot)$ represents the rank of the matrix and
$|\mathcal{A}|$ denotes the cardinality of set a $\mathcal{A}$.
$||\cdot||_F$ and $||\cdot||$ indicates \emph{Frobenius} and $\ell_2$- norms, respectively.

\section{System model}
\label{system}
\begin{figure}[!h]
\centerline{\psfig{figure=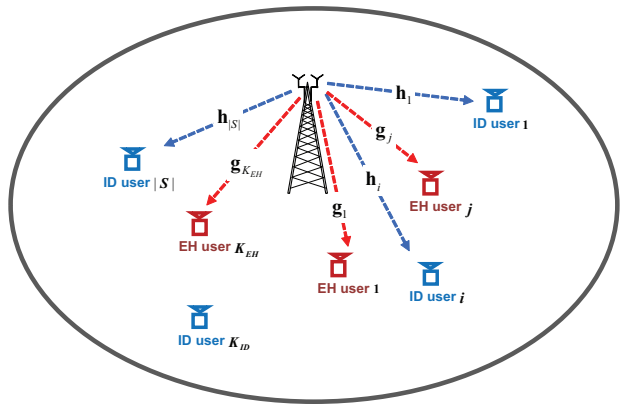,width=3.2in}}
\caption{System model for multi-user wireless information and power transfer}\label{fig:system}
\end{figure}

We consider a multiuser MISO downlink channel model for simultaneous wireless information and power transfer (SWIPT),
in which the BS has $M$ antennas, and ID and EH
users have one single antenna as shown in Fig. \ref{fig:system}. It is assumed that the total number of ID and EH users is $K_{ID}$ and $K_{EH}$, respectively. In a multiuser wireless information and power transfer (MU-WIPT) system, each ID user is served by its dedicated beamforming vector while EH users can harvest energy from all beamforming vectors. This paper focuses on a \emph{practical} joint beamforming vector design and its performance analysis under the assumption of perfect channel state information (CSI) knowledge for EH and ID users at BS.
The received signal of the $k^{th}$ ID user is given by
\begin{equation}
{y}_k=\textbf{h}_k\textbf{x}+{n}_k,~~~ k=1,...,|\mathcal{S}|
\label{eq:1}
\end{equation}
where $k$ indicates the $k^{th}$ user in a selected ID user set $\mathcal{S}$, $\textbf{h}_k\in \mathbb{C}^{1\times M}$ is the $k^{th}$
user's row channel vector, $\textbf{x}\in\mathbb{C}^{M\times 1}$ is
the transmit signal vector containing the information symbols of the
selected user set $\mathcal{S}$, ${n}_k$ is the independent complex Gaussian
noise of the $k^{th}$ user with unit variance. Here, the BS serves a subset of ID users, $\mathcal{S}\subseteq
\{1,...,K_{ID}\}$, with an average power constraint
$\mathbb{E}\{||\textbf{x}||^2\}=P$. Due to the limited
degrees of freedom at BS, the number of active users is restricted to
$|\mathcal{S}|\leq M$. The transmit signal can be written as
\begin{equation}
\textbf{x}=\sum_{i=1}^{|\mathcal{S}|}\textbf{w}_i x_i \label{eq:2}
\end{equation}
where $\textbf{w}_i\in\mathbb{C}^{M\times 1}$ is the arbitrary unit-norm
joint beamforming vector for the $i^{th}$ user, and $x_i$ is the $i^{th}$
user information symbol. The SINR for the $k^{th}$ ID user is given by
\begin{equation}
\mbox{SINR}_k=\frac{p_k | \textbf{h}_k \textbf{w}_k
|^{2}}{1+\sum_{i\neq k}p_i|{\textbf{h}_k
\textbf{w}_i|^2}}=\frac{p_k ||\textbf{h}_k||^2|\overline{\textbf{h}}_k \textbf{w}_k
|^{2}}{1+\sum_{i\neq k}p_i||\textbf{h}_k||^2|{\overline{\textbf{h}}_k
\textbf{w}_i|^2}}
\label{eq:3}
\end{equation}
where $p_i$ is the transmit power assigned to $i^{th}$ joint beamforming vector, $\overline{\textbf{h}}_k$ is the
$k^{th}$ ID user's channel direction information, i.e.,
$\textbf{h}_k/||\textbf{h}_k||$.

In the case of EH users, no baseband processing is needed to harvest the carried
energy through beamforming vectors \cite{Rui2011}. Based on the law of energy conservation, the harvested energy
is proportional to the total received power. The total energy harvested from all $|\emph{S}|$ beamforming vectors is then given by
\begin{equation}
\mbox{Q}=\sum_{i=1}^{|\mathcal{S}|}{\mbox{Q}_i}=\zeta\sum_{i=1}^{|\mathcal{S}|} p_i||\textbf{G}\textbf{w}_i||^2
\label{eq:4}
\end{equation}
where $\zeta$ is a constant that accounts for the loss in the energy transducer for converting from the harvested energy to
electrical energy and $\textbf{G}\in \mathbb{C}^{K_{EH}\times M}$ is the EH users' concatenated channel matrix. For convenience, $\zeta$ is assumed equal to $1$.

\subsection{Problem statement and review of the SWIPT state of the art}
\label{art}
In this paper, the objective of joint beamforming is to maximize the total energy harvested by all EH users while satisfying the ID users' SINR constraints. The optimization problem is given as
\begin{equation}
\max_{\{\textbf{w}_i,~i\in\emph{S}\}}\sum_{i\in\emph{S}}\mbox{Q}_i~~\mbox{s.t.}~~\mbox{SINR}_i \geq \gamma_i, \forall i~~\mbox{and} \sum_{i\in\emph{S}}p_i\leq P
\label{eq:opt}
\end{equation}
where $\gamma_i$ is the $i^{th}$ ID user's SINR constraint. This problem is a general non-convex QCQP. In \cite{Jie2013}, the optimal solution for a given $\emph{S}$ can be obtained by applying the SDR and using existing software, e.g., CVX \cite{CVX}, only if the feasibility of ID users' SINR constraints is guaranteed. Thus, the feasibility of this problem has to be verified by solving the following problem :
\begin{equation}
\mbox{find} \{\textbf{w}_i\},~i\in\emph{S}
~~\mbox{s.t.}~~\mbox{SINR}_i \geq \gamma_i, \forall i~~\mbox{and} \sum_{i\in\emph{S}}p_i\leq P.
\label{eq:opt_1}
\end{equation}
In the MU-WIPT system, it is difficult to find out the optimal solution taking into account the ID user selection and the feasibility of the SINR constraints simultaneously because it is not a convex problem.

According to \cite{CVX}, CVX supports Self-Dual-Minimization package (SeDuMi) solver for semidefinite programming (SDP). The asymptotic computational complexity of SDP in SeDuMi (including main and inner iterations) is given as $O(n^2m^{2.5}+m^{3.5})$ where $O$ is big $O$ notation to describe the
asymptotic computational complexity behavior, $n$ is the number of decision variables and $m$ is the number of rows of the linear matrix inequalities (LMIs) \cite{complexity}. Since $n=M|\emph{S}|$ and $m=M|\emph{S}|$, the total approximated complexity is $O(M^{4.5}|\emph{S}|^{4.5}+M^{3.5}|\emph{S}|^{3.5})$.
Although the beamforming vectors obtained from (\ref{eq:opt}) are the optimal solution for the given selected ID users,  the complexity that arises during the computation is too high to be utilized as a practical algorithm.
Furthermore, the solution of the optimization problem does not lend itself to insightful harvested energy and sum rate analysis. In this paper, we propose a \emph{practical} joint beamforming algorithm to maximize the harvested energy while maintaining relatively low complexity and compatibility with conventional MU-MIMO systems \cite{spencer2004}-\nocite{yoo2006}\cite{Lim2013}. In addition, the harvested energy and the sum rate achievable with the proposed joint beamforming algorithm are analyzed.

\subsection{MU-WIPT system model based on ZFBF}
\label{ZFBF}

Unlike conventional MU-MIMO systems, in the MU-WIPT system, one should deal with not only maximizing the sum ID rate, but also the total harvested energy. Initially, to maximize the sum rate with reduced complexity \cite{yoo2006}, let us consider ZFBF with semi-orthogonal user selection (SUS) for ID users. The detailed operation of SUS algorithm is summarized in Table \ref{table:algorithm} for convenience.

\begin{table}[!h]
\centering
\renewcommand\arraystretch{1.0}
\newcommand\whline{\noalign{\global\savedwidth\arrayrulewidth
                           \global\arrayrulewidth 2.5pt}%
                           \noalign{\global\arrayrulewidth\savedwidth}}
\caption{\fontsize{3.0mm}{3.0mm}semi-orthogonal user selection at the BS}
\begin{tabular}{l}
\fontsize{3.2mm}{3.2mm}
\textbf{Step 1 :} The first user from the initial user set
  $\mathcal{U}_1=\{1,\ldots,K_{ID}\}$.\\
  $\pi(1)=\displaystyle\arg\max_{k\in\mathcal{U}_1}||\textbf{h}_k||$\\
  \textbf{Step 2 :} $i^{th}$ user selection ($i=2$ for initiation)\\
  While $i\leq M$ \\
  $\mathcal{U}_i=\{1\leq k\leq K :
  |\overline{\textbf{h}}_k^H\overline{\textbf{h}}_{\pi(j)}|\leq\epsilon, 1\leq j\leq i\}$
  \\
  $\pi(i)=\displaystyle\arg\max_{k\in\mathcal{U}_i}\mbox{SINR}_k^{ZF}$,$~~~~~$$i=i+1.$\\
  end
  \\
  \textbf{Step 3 :} The best user set $\mathcal{S}=\{\pi(1),\ldots,\pi(|\mathcal{S}|)\}, |\mathcal{S}|\leq M$.\\
\end{tabular} \label{table:algorithm}
\end{table}

Since the BS has perfect knowledge of the CSI of ID users, the multiuser interference (MUI) is perfectly nullified via ZFBF.
To focus on the joint beamforming design, we assume that an equal power\footnote{In this paper, we investigate the beamforming design assuming equal power allocation as it is commonly assumed in conventional systems (LTE-A) \cite{bruno2013} and in the ZFBF literature \cite{yoo2007}\cite{jindal2008}. The additional performance gain of unequal power allocation over equal power allocation will nevertheless be discussed in Fig. \ref{fig:sim5} of Section \ref{simulation}.}
is allocated to each active ID user, i.e., $p_k=\rho=\frac{P}{|\mathcal{S}|}$.
In the ZFBF scheme, the received SINR  of (\ref{eq:3}) is reduced to
\begin{equation}
\mbox{SINR}_k^{ZF}=\rho ||\textbf{h}_k||^2|\overline{\textbf{h}}_k \textbf{w}^{ZF}_k|=\rho||\textbf{h}_k||^2\cos^2\theta_\epsilon\leq \rho||\textbf{h}_k||^2.
\label{eq:5}
\end{equation}
where $\theta_{\epsilon}=\cos^{-1}\sqrt{\frac{1-(M-1)\epsilon}{1-(M-2)\epsilon}(1+\epsilon)}$ in \cite{yoo2007}.
In the practical case of $K_{ID}\gg M$, BS should try to select mutually semi-orthogonal ID users (small $\epsilon$).
The inequality in (\ref{eq:5}) is tight when $\theta_{\epsilon}$ is small. When $\theta_{\epsilon}=0$, the received SINR can be interpreted as a tight upper bound. In this case, the normalized $k^{th}$ ID user's channel $\overline{\textbf{h}}_k^H$ is almost identical to the ZF beamforming vector of the $k^{th}$ ID user, i.e., $|\overline{\textbf{h}}_k\textbf{w}^{ZF}_k| = 1$. In the rest of paper, we assume that $K_{ID}\gg M$.
To characterize the SINR loss, we re-write the SINR constraints in (\ref{eq:opt}) as
\begin{equation}
\frac{\mbox{SINR}_i}{\mbox{SINR}_i^{ZF}} \geq {\mu_i}, \forall i
\label{eq:constraint}
\end{equation}
where $\mbox{SINR}_i^{ZF}$ is the received SINR achievable with ZFBF and $\mu_i$ is the target SINR ratio for the $i^{th}$ ID user. For $i^{th}$ joint beamforming vector or ID user, $1-\mu_i$ is then interpreted as a relative SINR loss, w.r.t. the SINR achievable with conventional MU-MISO ZFBF.
\section{Joint Beam-forming Design for MU-WIPT}
\label{joint}

In the proposed joint beamforming algorithm, BS selects a beamforming vector dedicated to a specific ID user and steers it to the optimal direction to maximize the total harvested energy while satisfying the SINR constraints of ID users selected by SUS.
In Section \ref{optimal_EH}, it is demonstrated that the optimal direction is the eigenvector corresponding to the largest eigenvalue of EH users' channel space, $\textbf{G}^H\textbf{G}$ in \emph{Theorem 1}.
Based on \emph{Theorem 1}, we describe the proposed iterative joint beamforming algorithm to maximize the total harvested energy and satisfy the SINR constraints of the selected ID users in Section \ref{algorithm}.

\subsection{Optimal direction to maximize the total harvested energy}
\label{optimal_EH}

The optimal direction to maximize the total harvested energy can be interpreted on the geometric space.
In \cite{son2010}, it was proven that the effective channel gains in the space limited by an $N \times M$
MIMO channel form a $\min(M,N)$-dimensional ellipsoid in the given channel space. The right singular vectors and singular values of the MIMO channel become axis and radius in the $\min(M ,N)$-dimensional ellipsoid, respectively.
In this section, we demonstrate that the harvested energy writes as a concave function of the direction of the beamforming vector. This function forms a $\min(K_{EH},M)$-dimensional ellipsoid in the given EH users' channel space, $\textbf{G}^H\textbf{G}$.

\emph{\textbf{Theorem} 1:} For a given arbitrary unit beamforming vector $\textbf{w}$, the harvested energy, $||\textbf{G}\textbf{w}||^2$, lies on a $\min(K_{EH},M)$-dimensional ellipsoid with axis $\textbf{v}_i$ and radius $\sqrt{\lambda_i}$, where $\textbf{v}_i$ and $\lambda_i$ are the $i^{th}$ eigenvector and eigenvalue of $\textbf{G}^H\textbf{G}$, respectively.

\begin{figure}
\begin{minipage}[!t]{3.2in}
\begin{center}{\psfig{figure=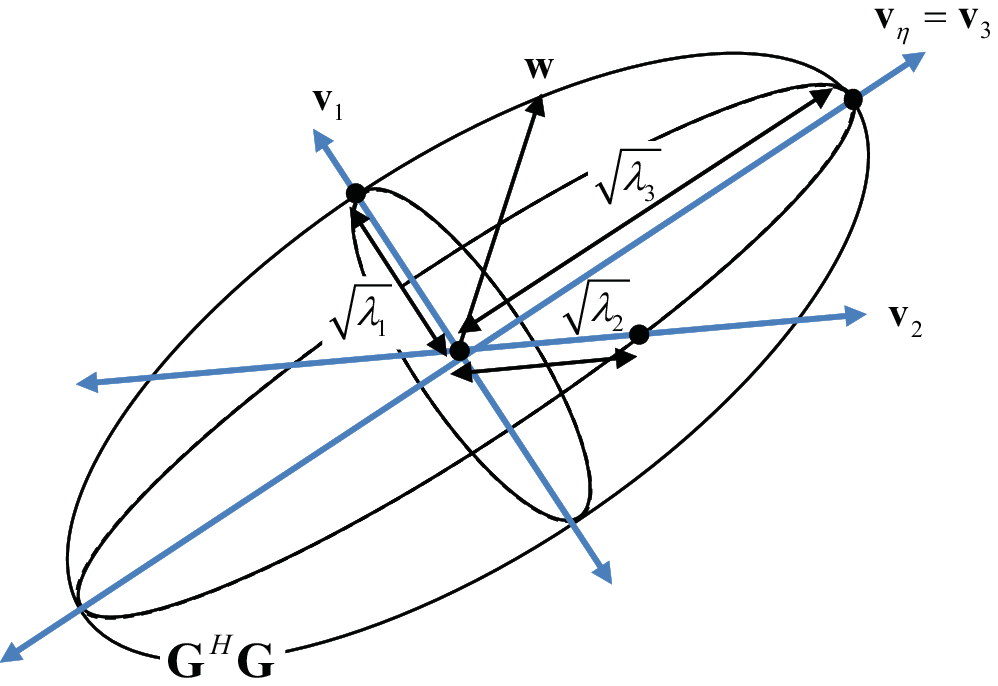,width=2.8in}}\end{center}
{\begin{center}(a) Example : 3-dimensional ellipsoid for EH rate when $\min(K_{EH},M)=3$
\end{center}}
\end{minipage}
\begin{minipage}[!t]{3.2in}
\vspace{0.2in}
\begin{center}{\psfig{figure=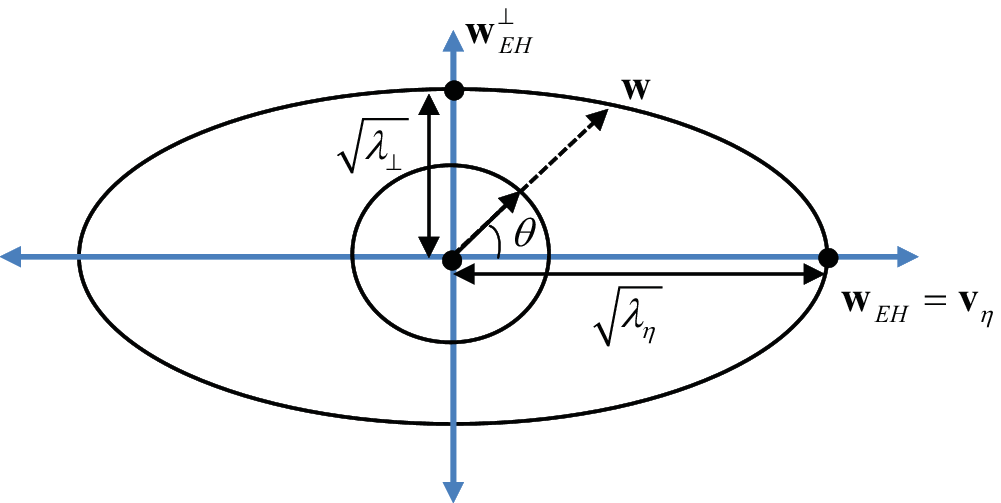,width=3.0in}}\end{center}
{\begin{center}(b) Geometric relation between $\textbf{w}_{EH}$ and $\textbf{w}$
\end{center}}
\end{minipage}
\caption{Representation of the ellipsoid in a geometric space of $\textbf{G}^H\textbf{G}$.}
\label{fig:1}
\end{figure}

\emph{\textbf{Proof}}:
Let $\textbf{G}^H\textbf{G}=\textbf{V}\mathbf{\Lambda}\textbf{V}^H$ using the eigenvalue decomposition, where
$\mathbf{\Lambda}=[0,\cdots,0,\lambda_1,\cdots,\lambda_\eta] \in\mathbb{C}^{M\times M}$ is the increasing-order diagonal matrix for eigenvalues, $\eta=\min(K_{EH},M)$ is the rank of $\textbf{G}^H\textbf{G}$ and $\textbf{V}=[\textbf{v}_i,\cdots,\textbf{v}_M] \in\mathbb{C}^{M\times M}$.
The energy harvested from $\textbf{w}$ can be represented as
\begin{equation}
\textsf{g}_{EH}(\textbf{w})=||\textbf{G}\textbf{w}||^2=\textbf{w}^H\textbf{G}^H\textbf{G}\textbf{w}
=\textbf{w}^H\textbf{V}\mathbf{\Lambda}\textbf{V}^H\textbf{w}
=\sum_{i=1}^{\eta}\lambda_i\textbf{w}^H\textbf{v}_i\textbf{v}_i^H\textbf{w}
=\sum_{i=1}^{\eta}\lambda_i\cos^2\theta_i.
\label{eq:6}
\end{equation}
Due to $\sum_{i=1}^{\eta}\cos^2\theta_i=1$, (\ref{eq:6}) is derived as
\begin{equation}
\sum_{i=1}^{\eta}\frac{\lambda_i\cos^2\theta_i}{\lambda_i}=\sum_{i=1}^{\eta}\frac{|\alpha_i|^2}{\lambda_i}=1
\label{eq:7}
\end{equation}
where $\alpha_i=\sqrt{\lambda_i}\textbf{w}^H\textbf{v}_i$ and $|\alpha_i|=\sqrt{\lambda_i}\cos\theta_i$.
Eq. (\ref{eq:7}) is an ellipsoid in which $\sqrt{\lambda_i}$ is a radius along the axis $\textbf{v}_i$
as shown in Fig. \ref{fig:1} (a). Therefore the coordinates of the harvested energy for an arbitrary unit beamforming vector $\textbf{w}$ are on the surface of the $\eta$-dimensional ellipsoid. $\blacksquare$

From \emph{Theorem 1}, the optimal beamforming vector $\textbf{w}_{EH}$ that maximizes the harvested energy in the space of $\textbf{G}^H\textbf{G}$ is the eigenvector corresponding to the largest eigenvalue of $\textbf{G}^H\textbf{G}$.
To measure the energy harvested by steering a given beamforming vector, let us define the normalized harvested energy as
\begin{equation}
\textsf{g}_{EH}(\textbf{w})=\textbf{w}^H\textbf{V}\mathbf{\Lambda}\textbf{V}^H\textbf{w}.
\label{normal}
\end{equation}
The given unit beamforming vector, $\textbf{w}$, is decomposed as
\begin{equation}
\textbf{w}=\cos\theta\textbf{w}_{EH} + \sin\theta\textbf{w}^{\bot}_{EH}
\label{eq:8}
\end{equation}
where
$\textbf{w}^{\bot}_{EH}=\frac{\textbf{w}-(\textbf{w}_{EH}^H\textbf{w})\textbf{w}_{EH}}
{||\textbf{w}-(\textbf{w}_{EH}^H\textbf{w})\textbf{w}_{EH}||}$.
Applying (\ref{eq:8}) to (\ref{normal}), the normalized harvested energy is derived as
\begin{eqnarray}
\textsf{g}_{EH}(\textbf{w})&=&(\cos\theta\textbf{w}_{EH} + \sin\theta\textbf{w}^{\bot}_{EH})^H\textbf{V}\mathbf{\Lambda}\textbf{V}^H(\cos\theta\textbf{w}_{EH} + \sin\theta\textbf{w}^{\bot}_{EH})\nonumber \\
&=&\textsf{g}_{EH}(\textbf{w}_{EH})\cos^2\theta + \textsf{g}_{EH}(\textbf{w}^{\bot}_{EH})\sin^2\theta\nonumber \\
&=&\lambda_{\eta}\cos^2\theta + \lambda_{\bot}\cos^2(\frac{\pi}{2}-\theta).
\label{eq:9}
\end{eqnarray}
Based on (\ref{eq:9}), the normalized harvested energy obtained from $\textbf{w}$ can be reduced
to a $2$-dimensional ellipsoid (with radius $\sqrt{\lambda_{\eta}}$ and $\sqrt{\lambda_{\bot}}$ along axis
$\textbf{v}_{\eta}$ and $\textbf{w}_{EH}^{\bot}$, respectively) as shown in Fig. \ref{fig:1} (b).
Since $\lambda_{\eta} \geq \lambda_{\bot}$ and the harvested energy is given as a concave function of the direction of an arbitrary unit beamforming vector $\textbf{w}$, the optimal way to maximize the harvested energy is to steer the beamforming vector along the geodesic line towards $\textbf{w}_{EH}$.
Thus, the joint beamforming vectors should be steered towards the direction of $\textbf{w}_{EH}$.

\subsection{Proposed algorithm}
\label{algorithm}

\begin{table*}[!t]
\centering
\renewcommand\arraystretch{1.0}
\newcommand\whline{\noalign{\global\savedwidth\arrayrulewidth
                           \global\arrayrulewidth 4.0pt}%
                           \noalign{\global\arrayrulewidth\savedwidth}}
\caption{\fontsize{3.0mm}{3.0mm}Joint Information and Energy Beamforming Algorithm}
\begin{tabular}{l}
\hline\hline
\fontsize{3.5mm}{3.5mm} \textbf{Step 1. Initialization}\\
$t=1$, $r=|\emph{S}|$\\
$\textbf{w}_i^t=\textbf{w}_i^{ZF}$ for $i\in\emph{S}$ \\
$\textbf{w}_{EH}^t=\textbf{w}_{EH}$\\
\fontsize{3.5mm}{3.5mm} \textbf{Step 2. Best beamforming vector selection and steering}\\
$\nabla EH_i^t=\frac{\textsf{g}_{EH}(\textbf{w}_{EH}^t)-\textsf{g}_{EH}(\textbf{w}_{i}^{t})}{\arccos(|(\textbf{w}_{EH}^t)^H\textbf{w}_i^{t}|)}$
where $i\in\emph{S}$\\
\textbf{If} $\nabla EH_i^t < 0, ~~\nabla EH_i^t=0$ for $i\in\emph{S}$~~\textbf{end}\\
$b=\arg\max_{i\in\emph{S}} \nabla EH_i^t$\\
\textbf{While} $\mbox{SINR}_i > \mu_i\cdot\mbox{SINR}_i^{ZF} (=\gamma_i)$ for $i\in\emph{S}$\\
~~~$\theta_b^{ID}=\theta_b^{ID}+\Delta D$ where $0\leq \theta_b^{ID}\leq \arccos(|(\textbf{w}_{EH}^t)^H\textbf{w}_b^t|)$ ~~~and $\Delta D$ is the unit angular distance\\
~~~$\textbf{w}_{b}^t=\cos\theta^{ID}_b\textbf{w}_{b}^{t-1}+\sin\theta^{ID}_b\textbf{w}_{b}^{\bot}$\\
~~~Update $\mbox{SINR}_i$ based on $\textbf{w}_{b}^t$ for $i\in\emph{S}$\\
~~~\textbf{If} $\mbox{SINR}_i \leq \gamma_i$ for any $i\in\emph{S}$\\
~~~$\emph{S}_{N}(r)=i$,~~$\textbf{w}_{b}^t=\cos(\theta^{ID}_b-\Delta D)\textbf{w}_{b}^{t-1}+\sin(\theta^{ID}_b-\Delta D)\textbf{w}_{b}^{\bot}$\\
~~~\textbf{end}\\
\textbf{end}\\
$\nabla EH_b^t=0$ and repeat {Step 2} until $\nabla EH_i^t=0$ for $\forall i$\\
\fontsize{3.5mm}{3.5mm} \textbf{Step 3. Optimal direction update}\\
\textbf{If} $r=0$,
~~Update $\textbf{w}_i^*=\textbf{w}_i^t$ for $i\in\emph{S}$ and Quit \\
\textbf{else}
~$r=r-1$, $t=t+1$, $\textbf{w}_{EH}^{t}=\mbox{eig}(\textbf{G}_N^H\textbf{G}_N)$ and repeat {Step 2} \\
\textbf{end}\\
\hline\hline
\end{tabular} \label{table:algorithm1}
\end{table*}

The proposed algorithm is divided into three parts, namely the gradient-based beamforming vector selection, the steering of the selected beamforming vector and the update of the optimal EH direction.
In each iteration, the BS selects the best beamforming vector among $\textbf{w}_{\forall i}$, and then steers it towards the optimal EH direction in order to maximize the total harvested energy and to guarantee the ID users' SINR constraints.
In other words, the selected beamforming vector is steered to maximize the total harvested energy at the cost of ID users' SINR loss because steering the beamforming vector reduces the desired channel gain and increases the interference to other ID users.

\textbf{Step 1) Initialization} :
As mentioned in Section \ref{ZFBF}, ID users are selected by SUS and the initial beamforming vectors dedicated to the selected ID users are constructed based on ZFBF (initialized as $\textbf{w}_i^{ZF}$). Let us define iteration index, $t$, and indicator, $r$, to show the number of available beamforming vectors that can be steered in the next iteration. Initially, $r$ is equal to $|\emph{S}|$ and the optimal EH direction is $\textbf{w}_{EH}$, which is the eigenvector corresponding to the largest eigenvalue of $\textbf{G}^H\textbf{G}$ as mentioned in \ref{optimal_EH}.

\textbf{Step 2) Best beamforming vector selection and steering} :

The optimal approach to design the joint beamforming is to maximize the harvested energy while minimizing the loss of SINR. If SINR is a concave function of the beam vectors, we can determine the optimal steering direction to maximize a gradient, which is defined as the increased harvested energy normalized by the SINR loss per unit angular distance. However, it is difficult to find the optimal joint beamforming vectors via vector steering mechanism with low complexity. Thus, we propose the EH gradient to select the best beamformer, which has the largest increasing harvested energy rate per unit angular distance. Under given SINR constraints, the best beamforming vector is selected among the co-scheduled $|\emph{S}|$ beams as the one that leads to the largest EH gradient. Let $\textbf{w}_{i}^t$ be the $i^{th}$ joint beamforming vector in the $t^{th}$ iteration of the proposed algorithm in Table \ref{table:algorithm1}.
The EH gradient for $\textbf{w}_{i}^t$ is defined as
\begin{equation}
\nabla EH_i=\frac{\textsf{g}_{EH}(\textbf{w}_{EH}^t)-\textsf{g}_{EH}(\textbf{w}_{i}^t)}{\arccos(|({\textbf{w}_{EH}^t})^H\textbf{w}_i^t|)}
\label{eq:9_1}
\end{equation}
where $\textsf{g}_{EH}(\textbf{w}_{EH}^t)-\textsf{g}_{EH}(\textbf{w}_{i}^t)$ is the harvested energy gain achieved by steering $\textbf{w}_{i}^t$ up to $\textbf{w}_{EH}^t$ and $\arccos(|({\textbf{w}_{EH}^t})^H\textbf{w}_i^t|)$ is the angular distance between $\textbf{w}_{EH}^t$ and $\textbf{w}_i^t$.
Based on (\ref{eq:9_1}),
the criterion to select the best beamforming vector is to determine the one with the largest harvested energy gain per unit angle distance at each iteration and is given by
\begin{equation}
b=\arg\max_{i\in\emph{S}} \nabla EH_i
\label{eq:criterion}
\end{equation}
where $b$ is the best beam index in each iteration.

In order to guarantee the ID users' constraints and maximize the total harvested energy, the selected beamforming vector is steered towards the direction of $\textbf{w}_{EH}^t$ as long as the ID users' SINR constraints are satisfied.
The selected beamforming vector can be decomposed as
\begin{equation}
\textbf{w}_b^{t}=\cos\theta_b^{ID}\textbf{w}_b^{t-1} + \sin\theta_b^{ID}\textbf{w}_b^{\perp},
\label{eq:decom}
\end{equation}
where $\textbf{w}^{\bot}_{b}=\frac{\textbf{w}_{EH}^t-({\textbf{w}_{b}^{t-1}}^H\textbf{w}_{EH}^t)\textbf{w}_{b}^{t-1}}
{||\textbf{w}_{EH}^t-({(\textbf{w}_{b}^{t-1})}^H\textbf{w}_{EH}^t)\textbf{w}_{b}^{t-1}||}$ and
$0\leq \theta_b^{ID}\leq \arccos(|({\textbf{w}_{EH}^t})^H\textbf{w}_b|)$.
The beamforming vector can then be steered on the geodesic line from $\textbf{w}_{b}^{t-1}$ to $\textbf{w}_{EH}^t$.
If there is any ID user who has lower SINR than its own SINR constraint,
the ID user index is included in the set of $\emph{S}_{N}(r)$, steering the selected beamforming vector stops and the EH gradient of $\textbf{w}_{i}$ is updated to zero so as not to be selected in the next iteration.

\textbf{Step 3) Optimal EH direction update} :
At the first iteration, $t=1$, the optimal direction for maximizing the harvested energy is $\textbf{w}_{EH}^1=\textbf{w}_{EH}$.
After all EH gradients for each joint beamforming vector become zero in Step 2, the beamforming vectors cannot be further steered towards $\textbf{w}_{EH}$ in the next iteration due to the SINR constraints. However, there may remain ID users with higher SINR than the target SINR. In order to maximize the total harvested energy as much as possible, the optimal direction $\textbf{w}_{EH}^t$ should be updated to guarantee no additional interference to the ID users who are at the boundary of their own SINR constraints. In this step, the iteration index $t$ and indicator $r$ for the number of available optimal EH direction are updated to $r=r-1$ and $t=t+1$, respectively, unless $r$ is equal to zero.
The set of ID users who are closed to their own SINR constraint is given as $\emph{S}_{N}(r)$ in Step 2.
To prevent any increase of the interference to the $p(r)^{th}$ ID user who is at the edge of the SINR constraint
in the next iteration, the optimal EH direction has to be updated as the eigenvector corresponding to the largest eigenvalue of $\textbf{G}_N^H\textbf{G}_N$ where $\textbf{G}_N$ is the projected version of $\textbf{G}$ onto the null space of the ID users obtained from Step 2, i.e., $i \in \emph{S}_{N}$. Let define $\textbf{H}_N=[\overline{\textbf{h}}_{\emph{S}_{N}(r)};\ldots;\overline{\textbf{h}}_{\emph{S}_{N}(|\emph{S}|)}]$.
The null space of $\textbf{H}_N$, $\textbf{H}_{null}$, is simply obtained by singular value decomposition (SVD)\footnote{The null space of $\textbf{H}_N$ can be obtained using Gram-Schmidt process or matrix decomposition methods. For simplicity, singular value decomposition is utilized.}  of $\textbf{H}_N$,
\begin{equation}
\textbf{H}_{null}=\textbf{U}_R[(|\emph{S}|-r+2):M]
\label{eq:nulls}
\end{equation}
where $\textbf{H}_N=\textbf{U}_L\mathbf{\Sigma}\textbf{U}_R^H$.
From (\ref{eq:nulls}), $\textbf{G}_N$ is given as
\begin{equation}
\textbf{G}_N=\textbf{G}\textbf{H}_{null}\textbf{H}_{null}^H.
\label{eq:nullG}
\end{equation}
Therefore, the optimal direction to maximize the EH rate in iteration $t+1$ is given as
the eigenvector corresponding to the largest eigenvalue of $\textbf{G}_N^H\textbf{G}_N$, i.e.,
\begin{equation}
\textbf{w}_{EH}^{t+1}=\mbox{eig}(\textbf{G}_N^H\textbf{G}_N),
\label{eq:EH_up}
\end{equation}
where $\mbox{eig}(\textbf{A})$ is the eigenvector corresponding to the largest eigenvalue of $\textbf{A}$.
From (\ref{eq:nulls}), the optimal EH direction in the proposed algorithm can be updated maximum $|\emph{S}|-1$ times.  The total number of optimal EH direction utilized in the proposed algorithm is then $|\emph{S}|$ including $\textbf{w}_{EH}$. Thus, $r$ can be interpreted as the number of available optimal EH direction not to incur the additional interference to the ID users who are at the boundary of their own SINR constraints.
Therefore, once $r$ is equal to $0$, the algorithm is completed because there is no direction left to improve the total harvested energy. The proposed joint beamforming algorithm is summarized in Table \ref{table:algorithm1}.

The asymptotic complexity of each function utilized in the proposed algorithm is given as $O(M^3)$ for eigenvalue decomposition and matrix inversion, $O(M)$ for the others \cite{Golub}. The total number of iteration for the main loop is equal or less than $\min(|\emph{S}|+1,M)$. In each main loop, the total number of iterations is equal or less than $|\emph{S}|$. Unless $K_{ID}$ is very large, $|S|+1$ is equal or less than $M$, i.e. $\min(|\emph{S}|+1,M)=|\emph{S}|+1$. Taking into account the number of iterations of the main and inner loops, the total asymptotic computational complexity is given as
\begin{equation}
O(M^3 + M^3|\emph{S}|(|\emph{S}|+1)+M|\emph{S}|(|\emph{S}|+1))
\approx O(M^3+M^3|\emph{S}|^2 + M^3|\emph{S}|) < O(M^3|\emph{S}|^3 ).
\label{eq:complex}
\end{equation}
Therefore, the asymptotic computational complexity is significantly lower than $O(M^{4.5}|\emph{S}|^{4.5}+M^{3.5}|\emph{S}|^{3.5})$ discussed in Section \ref{art}.
Furthermore, the eigenvalue decomposition can be replaced by
\begin{equation}
\textbf{w}_{EH}^t=\frac{\textbf{w}_{EH}^{t-1}-\overline{\textbf{h}}_b\textbf{w}_{EH}^{t-1}\overline{\textbf{h}}_b^H}{||\textbf{w}_{EH}^{t-1}-\overline{\textbf{h}}_b\textbf{w}_{EH}^{t-1}\overline{\textbf{h}}_b^H||}.
\label{eq:reduce}
\end{equation}
In this case, the complexity for updating $\textbf{w}_{EH}$, i.e. $ M^3|\emph{S}|(|\emph{S}|+1)$ is reduced to $ M|\emph{S}|(|\emph{S}|+1)$. The total asymptotic computational complexity in (\ref{eq:complex}) is given as
$O(M^3 + M|\emph{S}|(|\emph{S}|+1)+M|\emph{S}|(|\emph{S}|+1))\approx O(M^3 + M|\emph{S}|^2 + M|\emph{S}|) < O(M^3)$ due to $|\emph{S}|\leq M$. The performance of the proposed joint beamforming with reduced complexity is discussed in Fig. \ref{fig:sim5} of Section \ref{simulation}.

\section{Performance analysis for MU-WIPT system}
\label{analysis}
In this section, the total harvested energy and sum rate achieved with the proposed joint beamforming are derived as a function of the target SINR ratio in MU-WIPT system. The asymptotic total harvested energy and sum rate is then analyzed considering very large $K_{EH}$ and $K_{ID}$. To analyze the tradeoff relationship between total harvested energy and sum rate in the proposed joint beamforming vectors, we assume that ID users have the same target SINR ratio, $\mu$, as a constraint. Then $1-\mu$ can show the relative SINR loss (w.r.t. the maximum SINR
achievable with ZFBF).

\begin{figure}
\begin{minipage}[!t]{3.2in}
\begin{center}{\psfig{figure=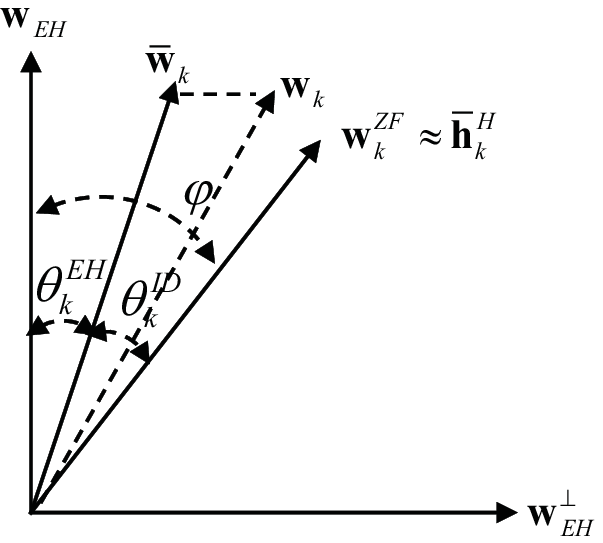,width=2.0in}}\end{center}
{\begin{center}(a)
\end{center}}
\end{minipage}
\begin{minipage}[!t]{3.2in}
\begin{center}{\psfig{figure=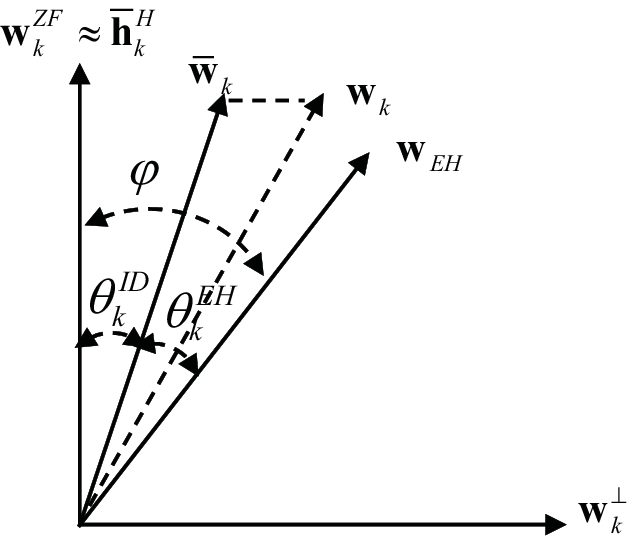,width=2.0in}}\end{center}
{\begin{center}(b)
\end{center}}
\end{minipage}
\caption{Representation of the geometric relation among $\textbf{w}^{ZF}_k$, $\textbf{w}_k$ and $\textbf{w}_{EH}$}
\label{fig:steer}
\end{figure}

\subsection{Sum rate analysis}
From \cite{yoo2006} and \cite{Karama2009}, the probability $\mbox{Pr}[k\in\mathcal{U}_{i+1}]$ in SUS is given as
\begin{equation}
\mbox{Pr}[k\in\mathcal{U}_{i+1}]=I_{\epsilon^2}(i,M-i)
\label{eq:pro}
\end{equation}
where
$I_x(\alpha_1,\alpha_2)=\sum_{j=\alpha_1}^{\alpha_1+\alpha_2-1}\left(\begin{array}{c}
                                          \alpha_1+\alpha_2-1 \\
                                          j
                                        \end{array}\right)x^j(1-x)^{\alpha_1+\alpha_2-1-j}$.
Based on the law of large numbers, the cardinality of $\mathcal{U}_{i}$, $|\mathcal{U}_{i}|$, is then approximated as
\begin{equation}
|\mathcal{U}_{i}|\approx K_{ID}I_{\epsilon^2}(i-1,M-i+1)
\label{eq:car}
\end{equation}
where $|\mathcal{U}_1|=K_{ID}$.
Note that $\pi(i)$-user is the best one selected from $\mathcal{U}_i$ in Table \ref{table:algorithm}, which has the maximum channel gain among $|\mathcal{U}_i|$ ID users.
Based on the order statistics \cite{david}, the probability density function (PDF) of $\pi(i)$ user's channel gain is given as
\begin{equation}
f_{\pi(i)}(x)=|\mathcal{U}_i|f_{\chi^2(2M)}(x)F_{\chi^2(2M)}(x)^{|\mathcal{U}_i|-1}
\label{eq:PDF}
\end{equation}
where $f_{\chi^2(2M)}(x)$ is PDF of Chi-square distribution with $2M$ degree of freedom, $f_{\chi^2(2M)}(x)=\frac{1}{\Gamma(M)}x^{M-1}e^{-x}$, $F_{\chi^2(2M)}(x)$ is its CDF, $F_{\chi^2(2M)}(x)=1-e^{-x}\sum_{p=0}^{M-1}\frac{1}{p!}x^p$, and $\Gamma(a)$ is the gamma function with parameter $a$, $\Gamma(a)=(a-1)!$ \cite{Proakis}\cite{Sharif2005}.

From (\ref{eq:5}) and (\ref{eq:constraint}), the SINR of $k^{th}$ ID user is given as
\begin{equation}
\mbox{SINR}_k\geq\mu\mbox{SINR}_k^{ZF}=\mu\rho||\textbf{h}_k||^2.
\label{eq:ID_SINR}
\end{equation}
Since each beamforming vector is steered as far as the ID users' SINR constraints are satisfied in the proposed algorithm, the lower bound given in (\ref{eq:ID_SINR}) is quite accurate, which is confirmed in Fig. \ref{fig:sim2} of Section \ref{simulation}.
By taking $\mbox{SINR}_k\approx\mu\rho||\textbf{h}_k||^2$, the expected sum rate is approximated as
\begin{equation}
\mathbb{E}[C_{ID}]\approx\sum_{i=1}^{|\emph{S}|}\int_0^{\infty}\log(1+\mu\rho x)f_{\pi(i)}(x)dx.
\label{eq:upper}
\end{equation}
From (\ref{eq:upper}), the sum rate loss is then derived as
\begin{eqnarray}
\Delta R&=&\sum_{k=1}^{|\emph{S}|}\mathbb{E}\left[\log(1+\rho||\textbf{h}_k||^2)-\log(1+\mu\rho||\textbf{h}_k||^2)\right]\nonumber \\
&=&\sum_{i=1}^{|\emph{S}|}\int_0^{\infty}\log\left(\frac{1+\rho x}{1+\mu\rho x}\right)f_{\pi(i)}(x)dx.
\label{eq:ID_loss_d}
\end{eqnarray}
At high SNR, the sum rate loss in (\ref{eq:ID_loss_d}) can be approximated as
\begin{equation}
\Delta R\approx-|\emph{S}|\log(\mu).
\label{eq:ID_loss_app}
\end{equation}
In (\ref{eq:ID_loss_app}), decreasing $\mu$ makes the ID rate loss proportional to $\log(\frac{1}{\mu})$.
Furthermore, since the increase of $K_{ID}$ leads to higher $|\emph{S}|$ during ID user scheduling,
$\Delta R$ is increased according to the increase of $K_{ID}$. If $|\emph{S}|=M$, the additional increase of $K_{ID}$ does not impact on $\Delta R$. However, the sum rate is still increased due to the increased SINR based on the order statistics as shown in (\ref{eq:PDF}).

\subsection{Harvested energy analysis}
\label{eh}
In Step 3 of Table \ref{table:algorithm1}, the joint beamforming vectors are steered to the direction of updated $\textbf{w}_{EH}$ for maximizing the total harvested energy. Thus, the joint beamforming vectors obtained from Table \ref{table:algorithm1} are not guaranteed to lie on the space spanned by $\textbf{w}^{ZF}_k$ and $\textbf{w}_{EH}$ due to (\ref{eq:decom}) and (\ref{eq:EH_up}) as shown in Fig. \ref{fig:steer} (a).
The projected $k^{th}$ joint beamforming vector onto the space spanned by $\textbf{w}^{ZF}_k$ and $\textbf{w}_{EH}$ is decomposed by
\begin{equation}
\overline{\textbf{w}}_k=\cos\theta_k^{EH}\textbf{w}_{EH} + \sin\theta_k^{EH}\textbf{w}_{EH}^{\bot}
\label{eq:14}
\end{equation}
where $\textbf{w}^{\bot}_{EH}=\frac{\overline{\textbf{w}}_k-(\textbf{w}_{EH}^H\overline{\textbf{w}}_k)\textbf{w}_{EH}}
{||\overline{\textbf{w}}_k-(\textbf{w}_{EH}^H\overline{\textbf{w}}_k)\textbf{w}_{EH}||}$.
Thus, the energy harvested from (\ref{eq:14}) is lower bounded\footnote{When the number of selected users, $|\emph{S}|$, is closed to $M$, the lower bound for harvested energy in (\ref{eq:15}) is very tight because the distance between each joint beamforming vector and the space consisted of $\textbf{w}_{EH}$ and $\textbf{w}^{ZF}_k$ can be ignorable. It can be confirmed from the simulation results in Section \ref{simulation}.}
by
\begin{eqnarray}
\rho\textsf{g}_{EH}(\textbf{w}_k)&=&
\rho\sum_{i=1}^{\eta}\lambda_i|\textbf{v}_i^H\textbf{w}_k|^2\geq
\rho\sum_{i=1}^{\eta}\lambda_i|\textbf{v}_i^H\overline{\textbf{w}}_k|^2 \nonumber \\
&=&\rho\sum_{i=1}^{\eta}\lambda_i|\textbf{v}_i^H(\cos\theta_k^{EH}\textbf{w}_{EH} + \sin\theta_k^{EH}\textbf{w}_{EH}^{\bot}) |^2\nonumber \\
&=&\rho\lambda_{\eta}\cos^2\theta_k^{EH}+\rho\sin^2\theta_k^{EH}\sum_{i=1,i\neq\eta}^{\eta}\lambda_i|\textbf{v}_i^H\textbf{w}_{EH}^{\bot}|^2 \nonumber \\
&=&\rho\lambda_{\eta}\cos^2\theta_k^{EH}+\rho\sin^2\theta_k^{EH}\beta(1,M-2)\sum_{i=1,i\neq\eta}^{\eta}\lambda_i
\label{eq:15}
\end{eqnarray}
where the unit vectors $\textbf{v}_{i}^H$ (for $i\neq\eta$) and $\textbf{w}_{EH}^{\bot}$ are independent and isotropically distributed on the $(M-1)$ dimensional hyperplane orthogonal to $\textbf{w}_{EH}$. Hence,
$|\textbf{v}_i^H\textbf{w}_{EH}^{\bot}|^2$ is beta-distributed with parameters $(1,M-2)$ and a mean value of $1/(M-1)$. Since $\theta_k^{EH}$ and $\theta_k^{ID}$ are independent of the ID users' channel gains under the fixed $\mu$, $\theta_k^{EH}$ and $\theta_k^{ID}$ for all joint beamforming vectors are independent of SUS. Although
those value are dependent on the directions of the eigenvectors of $\textbf{G}^H\textbf{G}$, $\textbf{G}$ is independently generated regardless of the initial beamforming vectors, i.e. ZFBF vectors. Thus, we can presume that the distributions of the angle (i.e., $\theta_k^{EH}$ or $\theta_k^{ID}$) for all joint beamforming vectors are statistically identical with each other. This was demonstrated through the simulations for various simulation parameters. 
By ignoring a joint beamforming index in (\ref{eq:15}), the lower bound of the expected energy harvested from a joint beamforming vector $\textbf{w}$ is derived as
\begin{eqnarray}
\rho\mathbb{E}\left[\textsf{g}_{EH}(\textbf{w})\right]&\geq&\rho\mathbb{E}\left[\lambda_{\eta}\cos^2\theta^{EH}+\sin^2\theta^{EH}\beta(1,M-2)\sum_{i=1,i\neq\eta}^{\eta}\lambda_i\right]\nonumber\\
&=&\rho\mathbb{E}[\lambda_{\eta}\cos^2\theta^{EH}]+\frac{\rho}{M-1}\mathbb{E}[\sin^2\theta^{EH}(||\textbf{G}||_F^2-\lambda_{\eta})]\nonumber\\
&=&\rho\mathbb{E}[\lambda_{\eta}]-\rho\frac{M\mathbb{E}[\lambda_{\eta}]-\mathbb{E}[||\textbf{G}||_F^2]}{M-1}\mathbb{E}[\sin^2\theta^{EH}]
\label{eq:16}
\end{eqnarray}
where
$\mathbb{E}[||\textbf{G}||_F^2]=\mathbb{E}[\chi^2(2MK_{EH})]=MK_{EH}$ and $\mathbb{E}[\lambda_{\eta}]$ is calculated using PDF of the largest eigenvalue of a uncorrelated central Wishart maxtrix (i.e., $\textbf{G}^H\textbf{G}$) given in \cite{whichart}.
Therefore, in (\ref{eq:16}), the expectation of $\sin^2\theta^{EH}$ is approximated\footnote{Since the angle $\theta^{ID}$ is obtained from the proposed algorithm, it is very difficult to find the distribution. However, it is verified that the variance of the angle $\theta^{ID}$ is very small, i.e., $Var(\cos\theta^{ID}) \ll 10^{-3}$ and $Var(\cos\theta^{ID}) \ll 10^{-3}$.
Since the angle can be interpreted as a deterministic value for given simulation parameters, the covariance between $\sin\theta^{ID}$ and $\cos\theta^{ID}$ is very small. Through the simulations, we demonstrated that $Cov(\sin\theta^{ID},\cos\theta^{ID}) \ll 10^{-3}$. Therefore, the approximation in (\ref{eq:beta}) can be obtained from $\mathbb{E}[\sin\theta^{ID}\cos\theta^{ID}]\approx\mathbb{E}[\sin\theta^{ID}]\mathbb{E}[\cos\theta^{ID}]$. In addition, we make the following approximations $\mathbb{E}[\sin\theta^{ID}]\approx\sqrt{\mathbb{E}[\sin^2\theta^{ID}]}$ and
$\mathbb{E}[\cos\theta^{ID}]\approx\sqrt{\mathbb{E}[\cos^2\theta^{ID}]}$ in (\ref{eq:f_low}).
}
as
\begin{eqnarray}
\mathbb{E}[\sin^2\theta^{EH}]&=&\mathbb{E}[\sin^2(\varphi-\theta^{ID})]=\mathbb{E}[(\sin\varphi\cos\theta^{ID}-\cos\varphi\sin\theta^{ID})^2]\nonumber \\&=&\mathbb{E}[\sin^2\varphi]\mathbb{E}[\cos^2\theta^{ID}]+\mathbb{E}[\cos^2\varphi]\mathbb{E}[\sin^2\theta^{ID}]-2\mathbb{E}[\sin\varphi\cos\varphi]\mathbb{E}[\sin\theta^{ID}\cos\theta^{ID}]\nonumber \\&\approx&\frac{M-1}{M}\mathbb{E}[\cos^2\theta^{ID}]+\frac{1}{M}\mathbb{E}[\sin^2\theta^{ID}]-2\mathbb{E}[\sin\varphi\cos\varphi]\mathbb{E}[\sin\theta^{ID}]\mathbb{E}[\cos\theta^{ID}] \label{eq:beta}
\end{eqnarray}
where $\sin^2\varphi$ is beta-distributed with parameters $(M-1,1)$ and
a mean value of $(M-1)/M$, and $\cos^2\varphi$ is beta-distributed with parameters $(1,M-1)$ and
a mean value of $1/M$.
In Appendix \ref{app1}, $\mathbb{E}[\cos^2\theta^{ID}]$ is derived as
\begin{equation}
\mathbb{E}[\cos^2\theta^{ID}]\approx g(\mu) =\frac{1+\rho\mathbb{E}[||\textbf{h}||^2]\frac{|\emph{S}|-1}{M-1}}{\frac{1}{\mu}+\rho\mathbb{E}[||\textbf{h}||^2]\frac{|\emph{S}|-1}{M-1}}
\label{eq:up}
\end{equation}
where $\mathbb{E}[||\textbf{h}||^2]$ is the expected value across the selected ID users,
\begin{equation}
{\mathbb{E}[||\textbf{h}||^2]}=\frac{1}{|\emph{S}|}\sum_{i=1}^{|\emph{S}|}\int_{0}^{\infty}xf_{\pi(i)}(x)dx~~\mbox{for }\pi(i)\in\emph{S}.
\label{eq:h}
\end{equation}
In Appendix \ref{app2}, the expected value of $\sin\varphi\cos\varphi$ is given as
\begin{equation}
\mathbb{E}[\sin\varphi\cos\varphi]=(M-1)B(3/2,M-0.5).
\label{eq:avg_sc}
\end{equation}
Applying (\ref{eq:up}) and (\ref{eq:avg_sc}) into (\ref{eq:beta}), the expected energy harvested from a joint beamforming vector is derived as
\begin{eqnarray}
\rho\mathbb{E}[\textsf{g}_{EH}(\textbf{w})]&\geq&
\rho\mathbb{E}[\lambda_{\eta}]-\rho(M\mathbb{E}[\lambda_{\eta}]-\mathbb{E}[||\textbf{G}||_F^2])\nonumber \\
&\cdot&\left(\frac{(M-2)g(\mu)+1}{M(M-1)}-2B(3/2,M-0.5)\sqrt{g(\mu)(1-g(\mu))}\right)
\label{eq:f_low}
\end{eqnarray}
where $\mathbb{E}[\sin\theta^{ID}]\approx\sqrt{\mathbb{E}[\sin^2\theta^{ID}]}=\sqrt{(1-g(\mu))}$ and
$\mathbb{E}[\cos\theta^{ID}]\approx\sqrt{\mathbb{E}[\cos^2\theta^{ID}]}=\sqrt{g(\mu)}$ (See footnote 6).
Therefore, the expected total harvested energy is given as
$\mathbb{E}[C_{EH}]\geq\rho|\emph{S}|\mathbb{E}[\textsf{g}_{EH}(\textbf{w}_k)]$.
Similarly with (\ref{eq:16}), the achievable expected energy harvested from $k^{th}$ ZF beamforming vector is given by
\begin{equation}
\rho\mathbb{E}[\textsf{g}_{EH}(\textbf{w}^{ZF})]=\rho\mathbb{E}[\lambda_{\eta}]-\rho\frac{M\mathbb{E}[\lambda_{\eta}]-\mathbb{E}[||\textbf{G}||_F^2]}{M-1}\mathbb{E}[\sin^2\varphi].
\label{eq:zf_eh}
\end{equation}
From (\ref{eq:16}) and (\ref{eq:zf_eh}), the average energy harvesting gain from joint beamforming is given as a function of $\mu$ :
\begin{eqnarray}
\Delta EH&=&\rho|\emph{S}|\frac{M\mathbb{E}[\lambda_{\eta}]-\mathbb{E}[||\textbf{G}||_F^2]}{M-1}(\mathbb{E}[\sin^2\varphi]-\mathbb{E}[\sin^2\theta^{EH}])\nonumber\\
&=&\rho|\emph{S}|(M\mathbb{E}[\lambda_{\eta}]-\mathbb{E}[||\textbf{G}||_F^2])\nonumber \\
&\cdot&\left(\frac{(M-2)(1-g(\mu))}{M(M-1)}+2B(3/2,M-0.5)\sqrt{g(\mu)(1-g(\mu))}\right).
\label{eq:del_eh}
\end{eqnarray}

\subsection{Asymptotic Performance Analysis}
\subsubsection{Sum Rate with $K_{ID}\rightarrow\infty$}
By borrowing the asymptotic sum rate analysis for MU-MISO ZFBF in \cite{yoo2006},
the lower bound of asymptotic sum rate for ID users is derived as
\begin{equation}
\mathbb{E}[C_{ID}]\sim M\log(1+\mu\rho\log K_{ID})=M\log(1+\rho\log {K_{ID}}^{\mu})
\label{eq:asym_ID2}
\end{equation}
where $x \sim y$ indicates that $\lim_{K_{ID}\rightarrow\infty}x/y=1$.
In addition, since the ID user's channel gain obtained from the proposed joint beamforming algorithm can be approximated as $\mu||\textbf{h}_k||^2$ as shown in (\ref{eq:ID_SINR}), the upper bound of the asymptotic sum rate is identical to (\ref{eq:asym_ID2}) based on \cite{masoud2007}.
In \cite{yoo2006}, it is shown that ZFBF with SUS in MU-MIMO system asymptotically has the same sum rate compared as that of DPC, namely it scales as $M\log(1+\rho\log K_{ID})$. Comparing with (\ref{eq:asym_ID2}), the sum rate loss of joint beamforming in the MU-WIPT system can be interpreted as originating from a reduced multiuser diversity gain. In other words, the decrease of $\mu$ incurs a loss of multiuser diversity gain in MU-WIPT system. In addition, the asymptotic rate loss between joint beamforming and ZFBF with SUS in MU-WIPT systems is given as
\begin{eqnarray}
\Delta R_{K_{ID}\rightarrow\infty}&=&\log(1+\rho\log K_{ID})-\log(1+\rho\log {K_{ID}}^\mu)\nonumber \\
&\approx&\log(\rho\log K_{ID})-\log(\rho\log {K_{ID}}^\mu)=\Delta R.
\label{eq:infR}
\end{eqnarray}
As shown in (\ref{eq:infR}), the rate loss is given as a constant $\Delta R$ even if $K_{ID}$ approaches infinity.

\subsubsection{Harvested Energy with $K_{EH}\rightarrow\infty$}

The normalized EH rate obtained from an arbitrary beamforming vector is derived as
\begin{eqnarray}
\lim_{K_{EH}\rightarrow\infty}\frac{\textsf{g}_{EH}(\textbf{w})}{K_{EH}}
&=&\lim_{K_{EH}\rightarrow\infty}\frac{||\textbf{G}\textbf{w}||^2}{K_{EH}}=\frac{1}{K_{EH}}\sum_{i=1}^{K_{EH}}|\textbf{g}_i\textbf{w}|^2\nonumber \\
&=&\frac{1}{K_{EH}}\sum_{i=1}^{K_{EH}}||\textbf{g}_i||^2|\overline{\textbf{g}}_i\textbf{w}|^2=\frac{1}{K_{EH}}\sum_{i=1}^{K_{EH}}\chi^2_{2M}\cdot\beta(1,M-1)\nonumber\\
&=&\frac{1}{K_{EH}}\sum_{i=1}^{K_{EH}}\chi^2_{2}=\mathbb{E}[\chi^2_{2}]=1.
\label{eq:sp2}
\end{eqnarray}
where the product $\chi^2_{2M}\beta(1,M-1)$ is $\chi^2_2$ in \cite{jindal2008}.
Therefore, when the number of EH users, $K_{EH}$, goes to infinity,
\begin{equation}
\lim_{K_{EH}\rightarrow\infty}\textsf{g}_{EH}(\textbf{w})=K_{EH}.
\label{eq:sp3}
\end{equation}
In (\ref{eq:sp3}), it implies that the normalized harvested energy $\textsf{g}_{EH}(\textbf{w})$ is obtained from a $M$-dimensional hypersphere with the radius $\sqrt{K_{EH}}$ when $K_{EH}$ goes to infinity.
In other words, each $\lambda$ approaches $K_{EH}$. Then the EH gain in (\ref{eq:del_eh}) approaches zero because $\lim_{K_{EH}\rightarrow\infty}M\lambda_{\eta}-||\textbf{G}||_F^2=0$. It is shown that all beamforming algorithms asymptotically (for large $K_{EH}$) allow the same amount of energy to be harvested.

\subsection{Discussion for dedicated beamforming for EH users}
\label{add_beam}
In Section \ref{system}, if the number of selected ID users, $|\emph{S}|$, is less than $M$, the $M-|\emph{S}|$ dedicated EH beamforming vectors can be constructed in addition to the joint beamforming vectors in Table \ref{table:algorithm1}. However, the additional dedicated EH beamforming vectors have to be designed not to create additional interference to the selected ID users.
As mentioned in Section \ref{optimal_EH}, the optimal beamforming vector to maximize the harvested energy is $\textbf{w}_{EH}$, which is the eigenvector of $\textbf{G}^H\textbf{G}$ corresponding to the largest eigenvalue \cite{Rui2011}\cite{Jie2013}. It also means that rank-1 is optimal for maximizing the total harvested energy. Thus, the optimal dedicated EH beamforming vector should be the singular vector corresponding to the largest singular value of the projected $\textbf{G}$ onto the null space of the selected ID users' channel matrix.

However, in Step 3 of Table \ref{table:algorithm1}, joint beamforming vectors are steered towards the direction of the null space of ID users' channel direction in order to increase the total harvested energy.
In other words, the direction of dedicated beamforming vector is identical with that of the updated $\textbf{w}_{EH}$ in the last iteration of Step 3.
Therefore, an additional beamforming vector for EH users cannot improve the total harvested energy in the proposed joint beamforming algorithm.

\subsection{Limited feedback based MU-SWIPT}
\label{feedback}
In this section, we consider a limited feedback model where
each ID and EH user feeds back to the BS $B_{ID}$ and $B_{EH}$ bits for the quantized channel direction information (CDI) based on random vector quantization (RVQ, \cite{Chun2007}) as well as a perfect channel magnitude in the form of a channel quality information (CQI)\footnote{The number of bits for quantizing CQI can be kept relatively small and the quantized CDI is a more critical factor that affects to the performance. To concentrate on the impact of quantized CDI, the perfect CQI is assumed in this paper.}. The proposed joint beamforming algorithm can then be performed based on the quantized channel information without any modification.
Since the sum rate of the selected ID users in the limited feedback model can be simply extended by making use in (\ref{eq:upper}) of the SINR distribution of the limited feedback based MU-MIMO (e.g. \cite{yoo2007}), this section focuses on the analysis of the harvested energy.

In \emph{Theorem 1}, the quantized EH users' channel matrix $\widehat{\textbf{G}}$ changes the direction of axes and the radius of the ellipsoid constructed by $\textbf{G}$. Define a expected distance between axes of $\textbf{G}$ and $\widehat{\textbf{G}}$ as
$\Delta_d=\mathbb{E}[|\textbf{v}_i^H\widehat{\textbf{v}}_i|^2]$
where $\textbf{v}_i$ is the axis of the space $\textbf{G}^H\textbf{G}$ corresponding to $\lambda_i$ and
$\widehat{\textbf{v}}_i$ is the axis of the space $\widehat{\textbf{G}}^H\widehat{\textbf{G}}$ corresponding to $\widehat{\lambda}_i$.
By the Parseval-Plancherel energy conservation \cite{mallat},
$\widehat{\textbf{v}}_i$ can be interpreted as the quantized version of $\textbf{v}_i$ using $B_{EH}$ bits. Based on the expected distance between a channel vector and its quantized one in \cite{Chun2007}, the expected distance between axes of $\textbf{G}$ and $\widehat{\textbf{G}}$ is then given as
\begin{equation}
\Delta_{d}=1-2^{B_{EH}}B(2^{B_{EH}},\frac{M}{M-1}).
\label{eq:delta_D1}
\end{equation}
Applying (\ref{eq:delta_D1}) to (\ref{eq:9}), the expected eigenvalue corresponding to the quantized axis $\widehat{\textbf{v}}_k$, $\mathbb{E}[\widehat{\lambda}_k]$ is derived as
\begin{eqnarray}
\mathbb{E}[\widehat{\lambda}_k]=\mathbb{E}[\textsf{g}_{EH}(\widehat{\textbf{v}}_k)]&=&\mathbb{E}[\lambda_{k}]\Delta_d + \mathbb{E}[\lambda_{\bot}](1-\Delta_d)\nonumber \\
&=&\mathbb{E}[\lambda_{k}]\Delta_d + \frac{\mathbb{E}[||\textbf{G}||_F^2]-\mathbb{E}[\lambda_{k}]}{M-1}(1-\Delta_d)
\label{eq:quan_lam}
\end{eqnarray}
where $\mathbb{E}[\lambda_{\bot}]=\mathbb{E}[\sum_{i\neq k}\lambda_{i}|\textbf{v}_i^H{\textbf{w}_{EH}^{\bot}}|^2]=\mathbb{E}[\sum_{i\neq k}\lambda_{i}\beta(1,M-2)]=\frac{\mathbb{E}[||\textbf{G}||_F^2]-\mathbb{E}[\lambda_{k}]}{M-1}$.
Since the harvested energy lies on a $\min(K_{EH},M)$-dimensional ellipsoid with axis $\widehat{\textbf{v}}_i$ and radius $\sqrt{\widehat{\lambda}_i}$, and $g(\mu)$ only depends on $\mu$, the lower bound of the expected energy harvested from a joint beamforming vector in the limited feedback model is derived as
\begin{equation}
\rho\mathbb{E}[\textsf{g}_{EH}(\textbf{w})]\geq
\rho\mathbb{E}[\widehat{\lambda}_{\eta}]-\rho(M\mathbb{E}[\widehat{\lambda}_{\eta}]-\mathbb{E}[||\widehat{\textbf{G}}||_F^2])f(\mu) \label{eq:FB_low}
\end{equation}
based on (\ref{eq:f_low}), where $\mathbb{E}[\widehat{\lambda}_{\eta}]=\mathbb{E}[\lambda_{\eta}]\Delta_d +\frac{\mathbb{E}[||\textbf{G}||_F^2]-\mathbb{E}[\lambda_{\eta}]}{M}(1-\Delta_d)$ in (\ref{eq:quan_lam}), $\mathbb{E}[||\widehat{\textbf{G}}||_F^2]=\mathbb{E}[||\textbf{G}||_F^2]$ and
$f(\mu)=\left(\frac{(M-2)g(\mu)+1}{M(M-1)}-2B(3/2,M-0.5)\sqrt{g(\mu)(1-g(\mu))}\right)$.
From (\ref{eq:f_low}) and (\ref{eq:FB_low}), the harvested energy loss due to quantization error is defined as
\begin{eqnarray}
\Delta Q&=& \rho\mathbb{E}[{\lambda}_{\eta}]-\rho(M\mathbb{E}[{\lambda}_{\eta}]-\mathbb{E}[||{\textbf{G}}||_F^2])f(\mu)-\rho\mathbb{E}[\widehat{\lambda}_{\eta}]+\rho(M\mathbb{E}[\widehat{\lambda}_{\eta}]-\mathbb{E}[||\widehat{\textbf{G}}||_F^2])f(\mu)\nonumber\\ &=&(\mathbb{E}[{\lambda}_{\eta}]-\mathbb{E}[\widehat{\lambda}_{\eta}])(\rho-\rho M f(\mu))\nonumber\\
&=&(1-\Delta_d)\left(\mathbb{E}[{\lambda}_{\eta}]-\frac{\mathbb{E}[||\textbf{G}||_F^2]-\mathbb{E}[\lambda_{k}]}{M}\right)\sim2^{B_{EH}}B(2^{B_{EH}},\frac{M}{M-1})
\label{eq:EH_loss_Q}
\end{eqnarray}
where $\sim$ denotes proportionality.
From (\ref{eq:EH_loss_Q}), it is observed that the harvested energy loss is a function of $B_{EH}$ and not of $B_{ID}$.
It means that the harvested energy is independent with the initial ZFBF vectors in Step 1 of the proposed algorithm because the ID users' CDI quantization error affects on the initial ZFBF design.
Since the initial ZFBF vectors can be interpreted as the randomly generated ones from the viewpoints of EH users, they leads to the same average harvested energy regardless of $B_{ID}$.
Therefore, when $B_{EH}$ approaches to infinity, the harvested energy loss will be zero :
\begin{eqnarray}
&&\lim_{B_{EH}\rightarrow\infty}\Delta Q=\lim_{B_{EH}\rightarrow\infty}2^{B_{EH}}B\left(2^{B_{EH}},\frac{M}{M-1}\right)
=\lim_{B_{EH}\rightarrow\infty}\frac{2^{B_{EH}}\Gamma(2^{B_{EH}})\Gamma(\frac{M}{M-1})}{2^{B_{EH}}\Gamma(2^{B_{EH}}+\frac{M}{M-1})}\nonumber\\
&&=\underbrace{\lim_{B_{EH}\rightarrow\infty}\frac{\Gamma(2^{B_{EH}}+1)e^{\frac{1}{M-1}\ln(2^{B_{EH}}+1)}}{\Gamma(2^{B_{EH}}+1+\frac{1}{M-1})}}_{=1}\underbrace{\lim_{B_{EH}\rightarrow\infty}\frac{\Gamma(\frac{M}{M-1})}{e^{\frac{1}{M-1}\ln(2^{B_{EH}}+1)}}}_{=0}=0
\label{eq:Q0}
\end{eqnarray}
based on $\displaystyle\lim_{x\rightarrow\infty}\frac{\Gamma(x)}{\Gamma(x+a)}e^{a\log_2^x}=1$ in \cite{gradshteyn}.

\section{Simulation results}
\label{simulation}
In this section, simulation and numerical results are provided in order to validate the performance of the proposed algorithm and its numerical analysis. It is assumed that the signal attenuation from the BS to each user is $70$ dB (i.e. each user is located at an equal distance from the BS) and the loss to convert the harvested energy to electrical energy is zero, i.e., $\zeta=1$. The fading channels for all users are randomly generated from i.i.d. Rayleigh fading.
The total transmit power, $P$, and  noise variance at each receiver, $\sigma^2$, are set to $1$ Watt (i.e., $30$ dBm), and $-50$dBm, respectively. For user scheduling, SUS algorithm with $\epsilon = 0.3$ is employed as described in Table \ref{table:algorithm} because the optimal range of $\epsilon $ is between $0.2$ and $0.4$ as demonstrated in \cite{yoo2006}.

\begin{figure}[!h]
\centerline{\psfig{figure=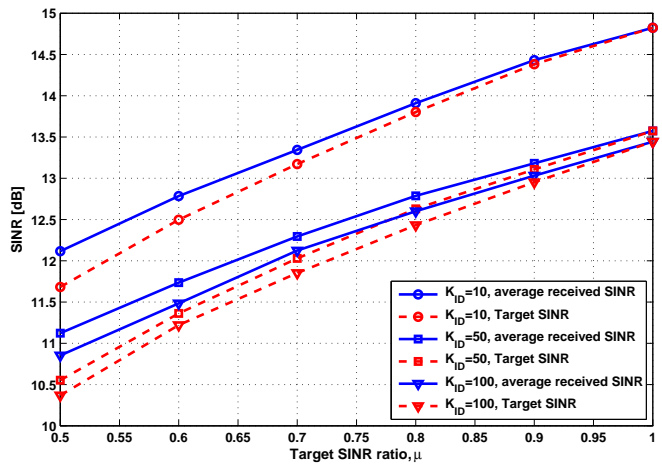,width=4.0in}}
\caption{Average target SINR and receive SINR according to $\mu$ when $K_{EH}=10$}\label{fig:sim2}
\end{figure}

\begin{figure}[!h]
\centerline{\psfig{figure=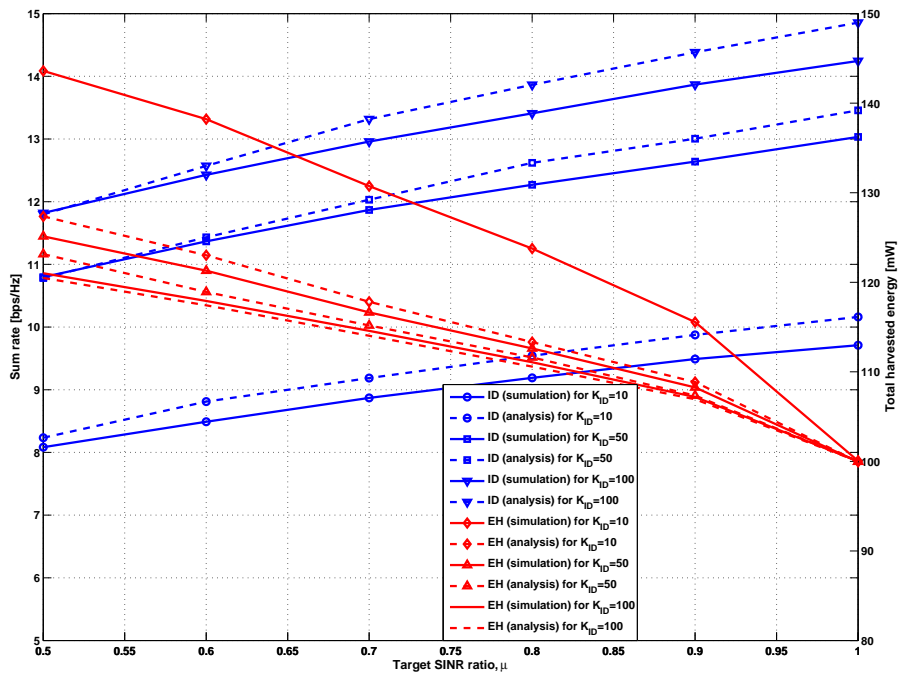,width=5.5in}}
\caption{Total harvested energy and sum rate performance comparison according to target SINR ratio, $\mu$ when $K_{EH}=10$}\label{fig:sim1}
\end{figure}

Figure \ref{fig:sim2} shows the average received and target SINRs of the proposed joint beamforming according to $\mu$. Since the received SINR is slightly higher than the target SINR, it confirms that the proposed joint beamforming algorithm satisfies the SINR constraints. Although the total harvested energy can be improved by further decreasing the gap between the target and received SINRs, the gap is consistently kept small (less than $0.5$ dB) as shown in Fig. \ref{fig:sim2}. Figure \ref{fig:sim1} shows the sum rate and total harvested energy according to $\mu$ and $K_{ID}$ when $K_{EH}$ is $10$. The average received and target SINRs for Fig. \ref{fig:sim1} is illustrated in Fig. \ref{fig:sim2}. All solid lines in Fig. \ref{fig:sim1} indicate the simulated results and all dotted lines shows the theoretical results based on (\ref{eq:upper}) and (\ref{eq:f_low}).

Since the total harvested energy is increased at the cost of the SINR loss, i.e., $(1-\mu)100$ $\%$, in the proposed joint beamforming algorithm, the increase of $\mu$ leads to the decrease of the total harvested energy and the increase of the sum rate as shown in Fig. \ref{fig:sim1}.
When $K_{ID}$ is increased, the EH rate is gradually decreased due to the increase of the number of joint beamforming vectors, $|\emph{S}|$, i.e., the number of co-scheduled ID users in SUS.
A low $|\emph{S}|$ has $M-|\emph{S}|$ dimensional free space into which the joint beamforming vectors can be steered to maximize the total harvested energy and not to increase multi-user interference. It means that joint beamforming vectors with a low $|\emph{S}|$ is closer to the optimal rank-1 beamforming vector $\textbf{w}_{EH}$ compared to joint beamforming vectors with a high $|S|$.
From this perspective, the increase of $|\emph{S}|$ degrades the harvested energy while it increases sum rate due to multiuser diversity.

The lower bounds based on (\ref{eq:f_low}) in Fig. \ref{fig:sim1} are quite tight and inline with their simulated results except for the case of $K_{ID}=10$.
The lower bound for the harvested energy is obtained from the projected joint beamforming vector onto the space spanned by $\textbf{w}_{EH}$ and $\textbf{w}^{ZF}$ as derived in (\ref{eq:14}) and (\ref{eq:15}).
Since the MU-WIPT system with a high $|\emph{S}|$ has a low (i.e., $M-|\emph{S}|$) dimensional free space, the distance between a joint beamforming vector $\textbf{w}_k$ and its projected version $\overline{\textbf{w}}_k$ becomes small. Conversely, a low $|\emph{S}|$ leads to the increase of the distance between $\textbf{w}_k$ and $\overline{\textbf{w}}_k$. Thus, the lower bounds at $K_{ID}=50$ and $100$ are tight compared to that of $K_{ID}=10$.

On the other hand, simulated sum rates are slightly upper-bounded by analytic sum rates due to the inequality in (\ref{eq:5}). However, the sum rate gap between them decreases as $\mu$ decreases. As shown in Fig. \ref{fig:sim2}, the average received SINR increases compared to the target SINR as $\mu$ decreases. Thus, the sum rate gap is compensated by SINR gap between received and target SINRs in Fig. \ref{fig:sim2}.

\begin{figure}[!h]
\centerline{\psfig{figure=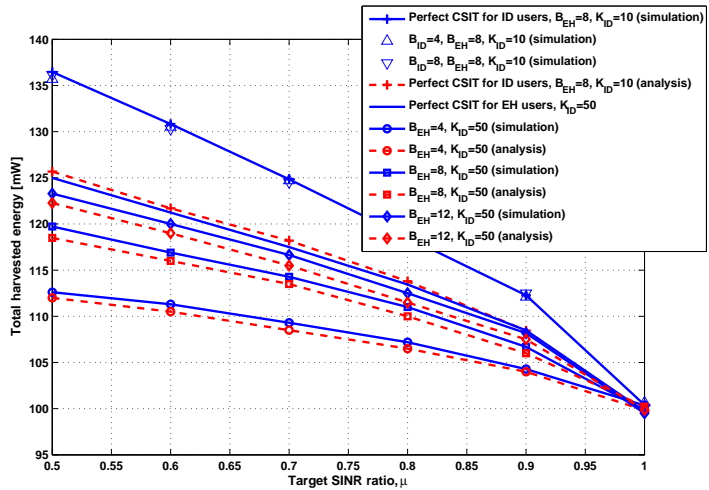,width=4.0in}}
\caption{Total harvested energy based on limited feedback when $K_{EH}=10$}\label{fig:sim4}
\end{figure}

Figure \ref{fig:sim4} shows the total harvested energy and its lower bound obtained from the limited feedback based joint beamforming. As shown in the simulated results with $K_{ID}=10$, there is no performance gap between them even if $B_{ID}$ is decreased. It shows that the ID user's quantization error cannot impact on the harvested energy in the proposed joint beamforming scheme as discussed in Section \ref{feedback}. Similarly to Fig. \ref{fig:sim4}, the lower bounds based on (\ref{eq:FB_low}) in Fig. \ref{fig:sim4} are quite tight except for the case of $K_{ID}=10$.
As shown in the simulated results with $K_{ID}=50$, the total harvested energy with limited feedback approaches that with perfect CSIT case as $B_{EH}$ is increased. This confirms (\ref{eq:Q0})


\begin{figure}[!h]
\centerline{\psfig{figure=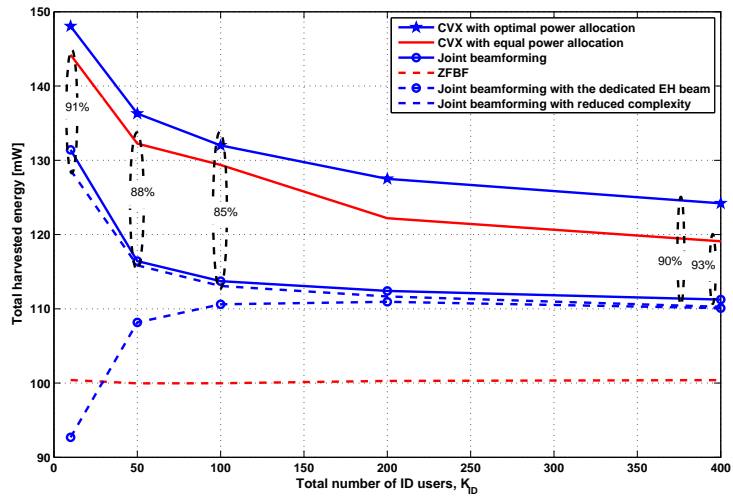,width=4.5in}}
\caption{Total harvested energy comparison according to $K_{ID}$ when $\mu=0.7$ and $K_{EH}=10$}\label{fig:sim5}
\end{figure}

Figure \ref{fig:sim5} shows the total harvested energy for beamforming schemes when $\mu=0.7$, $K_{EH}=10$
and $K_{ID}$ is in the range of $10$ to $400$.
In Fig. \ref{fig:sim5}, three beamforming schemes are utilized as the reference systems and SUS is adopted to select the ID users for all reference systems.
Although joint beamforming design in MU-MIMO system is not a convex problem as mentioned in Section \ref{art}, the optimal solution for the given selected ID users can still be numerically found.
In order to avoid the feasibility problem and to make fair comparison between the performance of the joint beamforming and CVX schemes, we modify the SINR constraints in (\ref{eq:opt}) as follows
\begin{equation}
\max_{\{\textbf{w}_i,~i\in\emph{S}\}}\sum_{i=1}^{|\mathcal{S}|} \rho||\textbf{G}\textbf{w}_i||^2
~~\mbox{s.t.}~~\mbox{SINR}_i \geq\mu\cdot\mbox{SINR}_i^{ZF}, \forall i.
\label{eq:mopt}
\end{equation}
The optimal solution for (\ref{eq:mopt}) can be obtained by CVX \cite{Jie2013}\cite{CVX}. Although it is difficult to utilize CVX as a practical algorithm due to its complexity, the performance obtained from CVX can be interpreted as the optimal performance for the selected ID users in MU-WIPT system. The second reference system is the conventional ZFBF method \cite{yoo2006}.
The third one is the modified joint beamforming algorithm which has the dedicated EH beam for EH users as described in Section \ref{add_beam}.

The proposed joint beamforming scheme shows better performance compared with ZFBF due to the beamforming vector steering to maximize the harvested energy. As discussed in Section \ref{add_beam}, the joint beamforming with the dedicated EH beam has a lower total harvested energy compared to the proposed joint beamforming. Since the increase of $K_{ID}$ leads to $|\emph{S}|\approx M$, the performance gap between joint beamforming schemes becomes very small. It means that there is no additional space to make the dedicated EH beam. 

The CVX scheme outperforms the proposed scheme because the optimal beamforming vectors are obtained by solving (\ref{eq:mopt}) for the given selected ID users.
However, the proposed joint beamforming scheme achieves a large percentage of the energy harvested from CVX scheme, i.e., $88$-$93\%$ for equal power allocation and $85$-$90\%$ even for optimal power allocation, as shown in Fig. \ref{fig:sim5}. Moreover, it is observed that the optimal power allocation makes a relatively small improvement compared to the equal power allocation in CVX. It is demonstrated that the joint beamforming vector design is more critical than the power allocation to increase the total harvested energy rather than an optimal power allocation in MU-SWIPT system.

In Fig. \ref{fig:sim5}, the joint beamforming with reduced complexity refers to the modified version based on (\ref{eq:reduce}). Although the replaced function, (\ref{eq:reduce}) cannot determine the optimal direction, it is demonstrated that the replaced function can provide a near-optimal EH direction and incurs a small performance loss compared to the joint beamforming with eigenvalue decomposition as shown in Fig. \ref{fig:sim5}.


\begin{figure}[!h]
\centerline{\psfig{figure=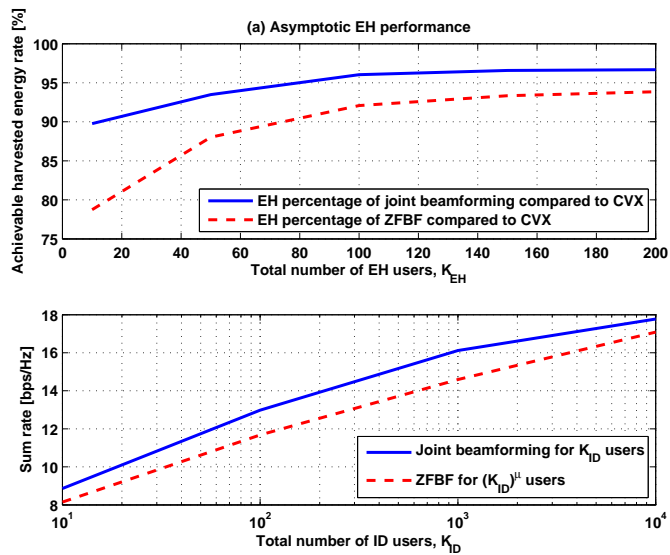,width=4.0in}}
\caption{Asymptotic EH and ID performance}\label{fig:sim6}
\end{figure}


Figure \ref{fig:sim6} (a) shows the harvested energy (i.e., percentage) normalized by the optimal performance obtained from CVX. As shown in Fig. \ref{fig:sim6}, the achievable harvested energy rate for the proposed beamforming scheme is increased from $90\%$ to $97\%$ as $K_{EH}$ is increased. The reason why the achievable harvested energy rate asymptotically goes to $100\%$ is that the energy harvested from any arbitrary beamforming vector approaches $K_{EH}$ at large $K_{EH}$ as seen from (\ref{eq:sp3}).
Thus, the harvested energy with ZFBF in Fig. \ref{fig:sim6} also increases as the $K_{EH}$ increases.


Figure \ref{fig:sim6} (b) shows the sum rate performance of both joint beamforming and ZFBF when the total number of ID users is respectively given as $K_{ID}$ and $(K_{ID})^\mu$.
From an asymptotic sum rate perspective, the proposed joint beamforming scheme exhibits the sum rate of a MU-MISO ZFBF with a reduced multiuser diversity gain as the number of ID users approaches infinity.
From (\ref{eq:asym_ID2}), it is observed that the sum ID rate of ZFBF for $(K_{ID})^\mu$ becomes the asymptotic lower-bound or the approximated sum rate of the proposed joint beamforming scheme. As shown in Fig. \ref{fig:sim6} (b), the sum rate of joint beamforming is lower bounded by that of ZFBF for $(K_{ID})^\mu$, and the gap between them is gradually decreased as $K_{ID}$ is increased.

\section{Conclusion}

For MU-WIPT system, we propose a joint beamforming algorithm to maximize the total harvested energy and to satisfy the SINR constraints taking into account the compatibility with the conventional MU-MIMO system.
When ID and EH users are simultaneously served by joint beamforming vectors, the harvested energy can be increased at the cost of an SINR loss of ID users. To characterize EH and ID performance, we analyze the EH and ID performance as a function of the target SINR ratio, $\mu$.
Through the simulation and numerical results, it is observed that the proposed joint beamforming scheme achieves a large percentage (about $85-93\%$) of the EH performance achievable with state of the art schemes but with a much lower complexity. In particular, it is shown that the number of joint beamforming vectors $|\emph{S}|$ and the target SINR ratio $\mu$ are the critical parameters that impact the performance tradeoff between sum rate and total harvested energy.
In the limited feedback based MU-SWIPT system, it is observed that the total harvested energy only depends on the quantization error of EH users, not on that of ID users.
Asymptotically (for large $K_{EH}$), it is shown that all beamforming algorithms allow the same amount of energy to be harvested because the harvested energy obtained from an arbitrary beamforming vector approaches $K_{EH}$. In addition, the proposed joint beamforming algorithm achieves a sum rate similar to that of a MU-MISO ZFBF with a reduced multiuser diversity gain as the number of ID users approaches infinity.

\appendices
\section{The proof of (\ref{eq:up})}
\label{app1}

As mentioned in Section \ref{eh}, the joint beamforming vectors obtained from Table \ref{table:algorithm1} are not guaranteed to lie on the space spanned by $\textbf{w}^{ZF}_k$ and $\textbf{w}_{EH}$ due to (\ref{eq:decom}) and (\ref{eq:EH_up}). However, the distance between the given space and each joint beamforming vector is marginal to affect the SINRs of ID users when $K_{ID} \gg M$.
By ignoring the distance between the given space and each joint beamforming vector, the geometrical relation among the joint beamforming vector of $k^{th}$ ID user, $\textbf{w}_k$, $\textbf{w}^{ZF}_k$, and $\textbf{w}_{EH}$ is constructed as shown in Fig. \ref{fig:steer} (a).
In (\ref{eq:3}), the desired signal gain is re-written as
\begin{equation}
\rho ||\textbf{h}_k||^2|\overline{\textbf{h}}_k \textbf{w}_k
|^{2}]=\rho ||\textbf{h}_k||^2\cos^2\theta_k^{ID}
\label{eq:11}
\end{equation}
and the interference signal gain is given as
\begin{eqnarray}
\rho\sum_{i\neq k}||\textbf{h}_k||^2|{\overline{\textbf{h}}_k
\textbf{w}_i|^2}&=&\rho\sum_{i\neq k}||\textbf{h}_k||^2|{{\textbf{w}_k^{ZF}}^H
\textbf{w}_i|^2}
=\rho||\textbf{h}_k||^2\sum_{i\neq k}{|{\textbf{w}_k^{ZF}}^H
(\cos\theta_i^{ID}\textbf{w}_i^{ZF} + \sin\theta_i^{ID}\textbf{w}_i^{\perp})|^2}\nonumber \\
&=&\rho||\textbf{h}_k||^2\beta(1,M-2)\sum_{i\neq k}\sin^2\theta_i^{ID}
\label{eq:12}
\end{eqnarray}
where $\textbf{w}^{\bot}_{i}=\frac{\textbf{w}_{EH}-({\textbf{w}_{i}^{ZF}}^H\textbf{w}_{EH})\textbf{w}_{i}^{ZF}}{||\textbf{w}_{EH}-({\textbf{w}_{i}^{ZF}}^H\textbf{w}_{EH})\textbf{w}_{i}^{ZF}||}$ and $\beta(\alpha_1,\alpha_2)$ is a beta-distributed random variable with parameter ($\alpha_1,\alpha_2$).
By taking (\ref{eq:5}), (\ref{eq:11}) and (\ref{eq:12}),
the received SINR ratio $\tau_k$ (cf. $\mu$ is a target SINR ratio) for the $k^{th}$ ID user (i.e., $k^{th}$ joint beamforming vector) is given as
\begin{equation}
\mu\approx\tau_k=\frac{\cos^2\theta_k^{ID}}{1+\rho||\textbf{h}_k||^2\beta(1,M-2)\sum_{i\neq k}\sin^2\theta_i^{ID}}.
\label{eq:tau}
\end{equation}
By ignoring a joint beamforming index of $\theta_k^{ID}$ in (\ref{eq:tau}), the expected SINR ratio is given as
\begin{eqnarray}
\mu&\approx&\mathbb{E}\left[\frac{1}{|\emph{S}|}\sum_{k=1}^{|\emph{S}|}\left\{\frac{\cos^2\theta_k^{ID}}{1+\rho||\textbf{h}_k||^2\beta(1,M-2)\sum_{i\neq k}\sin^2\theta_i^{ID}}\right\}\right]\nonumber\\
&=&\mathbb{E}\left[\frac{\cos^2\theta^{ID}}{1+\rho||\textbf{h}||^2\frac{1}{|\emph{S}|}\sum_{k=1}^{|\emph{S}|}\{\beta(1,M-2)\}(|\emph{S}|-1)\sin^2\theta^{ID}}\right]\label{eq:mu2}\nonumber\\
&=&\frac{\mathbb{E}[\cos^2\theta^{ID}]}{1+\rho\mathbb{E}||\textbf{h}||^2\frac{(|\emph{S}|-1)(1-\mathbb{E}[\cos^2\theta^{ID}])}{M-1}}
\label{eq:mu}
\end{eqnarray}
where $\textbf{h}=\frac{1}{|\emph{S}|}\sum_{k=1}^{|\emph{S}|}\textbf{h}_k$, $\beta(1,M-2)$ is a beta distributed random variable with parameter $(1,M-2)$ and a mean value of $1/M-1$, and $\mathbb{E}||\textbf{h}||^2$ is given in (\ref{eq:h}).
Since all instantaneous SINR ratio is approximated as $\mu$ during the derivation for (\ref{eq:mu}),
$\mu=\frac{1}{|\emph{S}|}\sum_{k=1}^{|\emph{S}|}\mu=\frac{1}{|\emph{S}|}\sum_{k=1}^{|\emph{S}|}A_k/\frac{1}{|\emph{S}|}\sum_{k=1}^{|\emph{S}|}B_k$ and $\mu=\mathbb{E}[\mu]=\mathbb{E}[A]/\mathbb{E}[B]$
due to $\frac{\mu}{|\emph{S}|}\sum_{k=1}^{|\emph{S}|}A_k=\frac{1}{|\emph{S}|}\sum_{k=1}^{|\emph{S}|}B_k$
and $\mu\mathbb{E}[A]=\mathbb{E}[B]$ where $A_k=\cos^2\theta_k^{ID}$, $B_k=1+\rho||\textbf{h}_k||^2\beta(1,M-2)\sum_{i\neq k}\sin^2\theta_i^{ID}$, $A=\frac{1}{|\emph{S}|}\sum_{k=1}^{|\emph{S}|}A_k$ and $B=\frac{1}{|\emph{S}|}\sum_{k=1}^{|\emph{S}|}B_k$.
After numerical manipulation of (\ref{eq:mu}), $\cos^2\theta^{ID}$ is derived as
\begin{equation}
\mathbb{E}[\cos^2\theta^{ID}]\approx g(\mu)=\frac{1+\rho\frac{|\emph{S}|-1}{M-1}\mathbb{E}||\textbf{h}||^2}{\frac{1}{\mu}+\rho\frac{|\emph{S}|-1}{M-1}\mathbb{E}||\textbf{h}||^2}
\label{eg:fun}
\end{equation}

\section{The proof of (\ref{eq:avg_sc})}
\label{app2}
Since the CDF of beta distribution is given as the regularized incomplete beta function,
the CDF for $\sin^2\varphi$ is derived as $F_{\sin^2\varphi}(x)=x^{M-1}$.
The PDF of $\sin\varphi$ is then given as $f_{\sin\varphi}(x)=2(M-1)x^{2M-3}$ because
the CDF of $\sin\varphi$ is given as
$F_{\sin\varphi}(x)=Pr[\sin^2\varphi \leq x^2]=x^{2(M-1)}$.
The expected value of $\sin\varphi\cos\varphi$ is then derived as
\begin{eqnarray}
\mathbb{E}[\sin\varphi\cos\varphi]&=&\mathbb{E}\left[\sin\varphi\sqrt{(1-\sin^2\varphi)}\right]
=\int_0^{1}x\sqrt(1-x^2)f_{\sin\varphi}(x)dx \nonumber \\
&=&2(M-1)\int_0^{1}\sqrt(1-x^2)x^{2(M-1)}dx=(M-1)B(3/2,M-0.5).
\label{eq:avg_sc1}
\end{eqnarray}

\bibliographystyle{ieeetr}
\bibliography{refs}

\end{document}